\documentclass[a4paper, 10pts]{article}
\usepackage{graphicx}
\usepackage{amssymb,amsmath}
\usepackage{mathptmx}
\usepackage[square, numbers]{natbib} 
\usepackage{array}
\usepackage{multirow} 
\usepackage{stmaryrd} 
\usepackage{hyperref} 

\hypersetup{
 	colorlinks=true,
 	linkcolor=black,
 	citecolor=black,
 	urlcolor=blue,
 	pdftitle={Continuous compliance: a proxy-based monitoring framework},
 	pdfauthor={Julien VEDANI, Fabien RAMAHAROBANDRO},
 	pdfsubject={continuous compliance, ORSA, life insurance, proxy, Least Squares Monte Carlo, Curve Fitting}
}

\newcommand{\cCite}[1]{\cite{#1}}

\newcolumntype{C}[1]{>{\centering} m{#1}}

\newlength{\tabularBigColWidth}
\newlength{\tabularSmallColWidth}
\graphicspath{{C:/Users/julien.vedani/Desktop/Articles/Article - conformite permanente/5 - Version LaTeX/}}

\newcommand{\footremember}[2]{%
   \footnote{#2}
    \newcounter{#1}
    \setcounter{#1}{\value{footnote}}%
}
\newcommand{\footrecall}[1]{%
    \footnotemark[\value{#1}]%
} 

\begin{document}

\title{Continuous compliance: a proxy-based monitoring framework}
\author{
    Julien VEDANI\footremember{milliman}{Milliman Paris, 14 rue Pergol\`ese, 75116 Paris, France}\footremember{uniLyon}{Universit\'e Claude Bernard Lyon 1, ISFA, 50 Avenue Tony Garnier, F-69007 Lyon, France}\footnote{Email: \url{julien.vedani@etu.univ-lyon1.fr}}
    \and Fabien RAMAHAROBANDRO\footrecall{milliman} \footnote{Email: \url{fabien.ramaharobandro@milliman.com}}
}

\date{December 17, 2013}
\maketitle

\renewcommand{\abstractname}{Abstract}
\begin{abstract}
Within the Own Risk and Solvency Assessment framework, the Solvency II directive introduces the need for insurance undertakings to have efficient tools enabling the companies to assess the continuous compliance with regulatory solvency requirements. Because of the great operational complexity resulting from each complete evaluation of the Solvency Ratio, this monitoring is often complicated to implement in practice. This issue is particularly important for life insurance companies due to the high complexity to project life insurance liabilities. It appears relevant in such a context to use parametric tools, such as Curve Fitting and Least Squares Monte Carlo in order to estimate, on a regular basis, the impact on the economic own funds and on the regulatory capital of the company of any change over time of its underlying risk factors.

In this article, we first outline the principles of the continuous compliance requirement then we propose and implement a possible monitoring tool enabling to approximate the eligible elements and the regulatory capital over time. In a final section we compare the use of the Curve Fitting and the Least Squares Monte Carlo methodologies in a standard empirical finite sample framework, and stress adapted advices for future proxies users.

\end{abstract}


\clearpage
\setlength{\parskip}{1ex}
\renewcommand{\thefootnote}{\arabic{footnote}}

\section{Introduction}

The Solvency II directive (European Directive 2009/138/EC), through the Own Risk and Solvency Assessment process, introduces the necessity for an insurance undertaking to be capable of assessing its regulatory solvency on a continuous yearly basis. This continuous compliance requirement is a crucial issue for insurers especially for life insurance companies. Indeed, due to the various asset-liability interactions and to the granularity of the insured profiles (see \emph{e.g.} Tosetti \emph{et al.} \cCite{Tosetti2003} and Petauton \cCite{Petauton2002}), the highly-stochastic projections of life insurance liabilities constitute a tricky framework for the implementation of this requirement.
 
In the banking industry the notion of continuous solvency has already been investigated through credit risk management and credit risk derivatives valuation, considering an underlying credit model (see e.g. Jarrow \emph{et al.} \cCite{Jarrow1997} and Lonstaff \emph{et al.} \cCite{Longstaff2005}). The notions of ruin and solvency are different in the insurance industry, due in particular to structural differences and to the specific Solvency II definitions. In a continuous time scheme these have been studied in a non-life ruin theory framework, based on the extensions of the Cramér-Lundberg model \cCite{Lundberg1903}, see \emph{e.g.} Pentikäinen \cCite{Pentikainen1982}, Pentikäinen \emph{et al.} \cCite{Pentikainen1989} and Loisel and Gerber \cCite{Loisel2012}. In a life insurance framework, considering more empirical schemes, closed formulas can be found under strong model assumptions. This field has for example been investigated in Bonnin \emph{et al.} \cCite{Bonnin2012} or Vedani and Virepinte \cCite{Vedani2011}. However, all these approaches are based on relatively strong model assumptions. Moreover, on a continuous basis the use of such approaches generally faces the problem of parameters monitoring and needs adaptations to be extended to the continuous compliance framework. 

Monitoring a life insurance liabilities is very complex and will have to introduce several stability assumptions in order to develop a practical solution. The great time and algorithmic complexity to assess the exact value of the Solvency Ratio of an insurance undertaking is another great issue. In practice, an only complete solvency assessment is required by the directive: the insurance undertakings have to implement a complete calculation of their Solvency Capital Requirement and of their eligible own funds at the end of the accounting year. We have identified two possibilities to investigate in order to implement a continuous compliance tool, either to propose a proxy of the Solvency Ratio, easy enough to monitor, or directly to address the solvency state (and not the solvency level). This last possibility leading to little information in terms of risk measurement we have chosen to consider the first one, based on the actual knowledge on the polynomial proxies applied to life insurance Net Asset Value (see \emph{e.g.} Devineau and Chauvigny \cCite{Devineau2011}) and Solvency Ratios (Vedani and Devineau \cCite{Vedani2013}), that is to say Least Squares Monte Carlo and Curve Fitting.

Throughout Section \ref{SectionIssuesAndOverview} we lay the foundations of the continuous compliance requirement adapted to life insurance. We underline and discuss the article designing the continuous compliance framework and present the major difficulties to address when implementing a monitoring tool. In Section \ref{QuantitativeApproach} we propose a continuous compliance assessment scheme based on a general polynomial proxy methodology. This tool is implemented in Section \ref{LSMC_ContinuousCompliance}, using a Least Squares Monte Carlo approach, on a standard life insurance product. The Least Squares Monte Carlo approach is generally preferred, in practice, to Curve Fitting because of its ``supposed'' advantages as soon as a large dimension context is concerned, which is the case in our continuous compliance monitoring scheme. We challenge this hypotheses in Section \ref{LSMCandCF_Comparison} where we implement both methodologies in various dimension frameworks and compared the obtained results.

\section{Continuous compliance}
\label{SectionIssuesAndOverview}

The requirement for continuous compliance is introduced in Article 45(1)(b) of the Solvency II Directive \cCite{Directive2009}: ``As part of its risk-management system every insurance undertaking and reinsurance undertaking shall conduct its own risk and solvency assessment. That assessment shall include at least the following: (...) the compliance, on a continuous basis, with the capital requirements, as laid down in Chapter VI, Sections 4 and 5''\footnote{Article 45(1)(b) also introduces continuous compliance ``with the requirements regarding technical provisions, as laid down in Chapter VI, Section 2''. This means that the companies should at all times hold technical provisions valued on the Solvency II basis. This implies that they have to be able to monitor the evolution of their technical provisions between two full calculations. The scope of this article is limited to continuous compliance with capital requirements.}.

In this section, we will first remind briefly what these capital requirements are and what they imply in terms of modelling and calculation. We will then discuss continuous compliance, what it entails and what issues it brings up for (re)insurance companies. Finally we will highlight some key elements to the setting of a continuous compliance framework in this business area.

\subsection{Capital requirements}

\subsubsection{Regulatory framework}

The capital requirements laid down in Chapter VI, Sections 4 and 5 are related to the Solvency Capital Requirement, or $SCR$ (Section 4), and the Minimum Capital Requirement, or $MCR$ (Section 5).

The $SCR$ corresponds to the Value-at-Risk of the basic own funds of the company subject to a confidence level of 99.5\% over a one-year period. It has to be calculated and communicated to the supervisory authority. Additionally, companies falling within the scope of the Financial Stability Reporting will have to perform a quarterly calculation (limited to a \emph{best effort basis}) and to report its results. Companies will have to hold eligible own funds higher or equal to the $SCR$. Failing to do so will trigger a supervisory process aiming at recovering a situation where the eligible own funds are in excess of the $SCR$. The $SCR$ can be calculated using the Standard Formula - a set of methodological rules set out in the regulatory texts - or an internal model (see below for further details).

The $MCR$ is a lower requirement than the $SCR$ , calculated and reported quarterly. It can be seen as an emergency floor. A breach of the $MCR$ will trigger a supervisory process that will be more severe than in the case of a breach of the $SCR$ and could lead to the withdrawal of authorization. The $MCR$ is calculated through a factor-based formula. The factors apply to the technical provisions and the written premiums in non-life and to the technical provisions and the capital at risk for life business. It is subject to an absolute floor and a floor based on the $SCR$. It is capped at 45\% of the $SCR$.

This paper focuses on the estimation of the eligible own funds and the $SCR$. Basically, the $MCR$ will not be used as much as the $SCR$ when it comes to risk management, and compliance with the $SCR$ will imply compliance with the $MCR$.

\subsubsection{Implementation for a life company}
\label{Regulatory_Implementation}

The estimation of the eligible own funds and the $SCR$ requires to carry out calculations that can be quite heavy. Their complexity depends on the complexity of the company's portfolio and the modeling choices that are made, in particular between the Standard Formula and an internal model. In this section, we present the key issues to be dealt with by a life insurer.

\paragraph{Implementation scheme.}

To assess the $SCR$ it is necessary to project economic balance sheets and calculate best estimates.

For many companies, the bulk of the balance sheet valuation lies in the estimation of these best estimates. This can imply quite a long and heavy process, since the assessment is carried out through simulations and is subject, amongst other things, to the following constraints:
\begin{itemize}
	\item updating the assets and liabilities model points;
	\item constructing a set of economic scenarios under the risk-neutral probability and checking its market-consistency;
	\item calibrating and validating the stochastic model through a series of tests (\emph{e.g.}: leakage test);
	\item running simulations.
\end{itemize}

The valuation of the financial assets may also be quite time-consuming if a significant part of the portfolio has to be marked to model.

\paragraph{$SCR$ calculation through the Standard Formula.} The calculation of the $SCR$ through the Standard Formula is based on the following steps:
\begin{itemize}
	\item calculation of the various standalone $SCR$;
	\item aggregation;
	\item adjustment for the risk absorbing effect of technical provisions and deferred taxes;
	\item calculating and adding up the capital charge for operational risk.
\end{itemize}

Each standalone $SCR$ corresponds to a risk factor and is defined as the difference between the current value of the eligible own funds and their value after a pre-defined shock on the risk factor. As a consequence, for the calculation of each standalone $SCR$ a balance sheet valuation needs to be carried out, which means that a set of simulations has to be run and that the assets must be valued in the economic conditions after shock.

\paragraph{$SCR$ calculation with a stochastic internal model.}

An internal model is a model designed by the company to reflect its risk profile more accurately than the Standard Formula. Companies deciding not to use the Standard Formula have the choice between a full internal model and a partial internal model. The latter is a model where the capital charge for some of the risks is calculated through the Standard Formula while the charge for the other risks is calculated with an entity-specific model.
There are two main categories of internal models\footnote{These approaches can be mixed within one model.}:
\begin{itemize}
	\item models based on approaches similar to that of the Standard Formula, whereby capital charges are calculated on the basis of shocks; the methodology followed in this case is the same as the one described in Subsection \ref{Regulatory_Implementation};
	\item 	fully stochastic models: the purpose of this type of model is to exhibit a probability distribution of the own funds at the end of a 1-year period, in order to subsequently derive the $SCR$ , by calculating the difference between the 99.5\% quantile and the initial value.
\end{itemize}

In the latter case, the calculations are based on a methodology called Nested Simulations. It is based on a twofold process of simulations:
\begin{itemize}
	\item	real-world simulations of the risk factors' evolution over 1 year are carried out;
	\item for each real-world simulation, the balance sheet must be valued at the end of the 1-year period. As per the Solvency II requirements, this valuation has to be market-consistent. It is carried out through simulations under the risk-neutral probability.
\end{itemize}

More details on Nested Simulations can be found in Broadie \emph{et al.} \cCite{Broadie2011} or Devineau and Loisel \cCite{Devineau2009b}.

\subsection{An approach to continuous compliance}

In the rest of this article we focus the scope of our study to life insurance.

\subsubsection{Defining an approach}
As mentioned above, the Solvency II Directive requires companies to permanently cover their $SCR$ and $MCR$. This is what we refer to as “continuous compliance” in this paper. The regulatory texts do not impose any specific methodology. Moreover the assessment of continuous compliance is introduced as an element of the Own Risk and Solvency Assessment (“ORSA”), which suggests that the approach is for each company to define.

Different approaches can be envisaged. Here below we present some assessment methodologies that companies can rely on and may combine in a continuous compliance framework.
\begin{itemize}
	\item \textbf{Full calculations}: \emph{i.e.} the same calculations as those carried out for annual reporting to the supervisory authority: this type of calculations can be performed several times during the year. However the process can be heavy and time-consuming, as can be seen from the description made in Subsection \ref{Regulatory_Implementation}. As a consequence, it seems operationally difficult to carry out such calculations more than quarterly (actually most companies are likely to run full calculations only once or twice a year).
	\item \textbf{“Simplified” full calculations}: companies may decide to run calculations similar to those described in the previous item but to freeze some elements. For example they could decide not to update the liabilities model points if the portfolio is stable and if the time elapsed since the last update is short; they could also decide to freeze some modules or sub-modules that are not expected to vary significantly over a short period of time.
	\item \textbf{Proxies}: companies may develop methods to calculate approximate values of their Solvency Ratio\footnote{Solvency Ratio = Eligible Own Funds / $SCR$.} ($SR$). Possible approaches include, among others, abacuses and parametric proxies.
	\item \textbf{Indicators monitoring}: as part of their risk management, companies will monitor risk indicators and set limits to them. These limits may be set so that respecting them ensures that some $SCR$ modules stay within a given range.
\end{itemize}

\subsubsection{Overview of the proposed approach}
\label{OverviewApproach}

The approach presented in this paper relies on the calibration of proxies allowing to estimate the $SR$ quickly and taking as input a limited number of easy-to-produce indicators. It has been developed for life companies using the Standard Formula.

\paragraph{Proxies: generic principles.}

Simplifying the calculations requires limiting the number of risks factors that will be monitored and taken into account in the assessment to the most significant. For most life insurance companies, these risk factors will be financial (\emph{e.g.}: stock level, yield curve).

In the framework described in the following sections, the proxies are supposed to be potentially used to calculate the $SR$ at any point in time. For operational practicality, the inputs have to be easily available. In particular, for each risk factor, an indicator will be selected for monitoring purpose and to be used as input for the proxy (see Section \ref{QuantitativeApproach} for more insight about proxies). The selected indicators will have to be easily obtainable and reflect the company's risk profile.

As explained in Section \ref{QuantitativeApproach}, our approach relies on the development and the calibration of proxies in order to calculate in a quick and simple way the company's Net Asset Value ($NAV$) and the most significant $SCR$ sub-modules. The overall $SCR$ is then calculated through an aggregation process based on the Standard Formula's structure and using the tools the company uses for its regulatory calculations. As a consequence, a selection has to be made regarding the sub-modules that will be calculated by proxy. The others are frozen or updated proportionally to a volume measure (\emph{e.g.} mortality $SCR$ set proportional to the technical provisions).

\paragraph{Continuous compliance framework.} Under Solvency II, companies will set a frequency (at least annual) for the full calculation of the $SCR$ \footnote{We are referring here to full calculations in the broad sense: the infra-annual calculations may be “simplified” full calculations.}. Additionally, they will set a list of pre-defined events and circumstances that will trigger a full calculation whenever they happen. The proxies will be used to estimate the $SR$ between two full calculations and should be calibrated every time a full calculation is performed. This process is summarized in Figure \ref{Figure1} below. 

\begin{figure}[h!]
\centering
\includegraphics[width=0.9\textwidth]{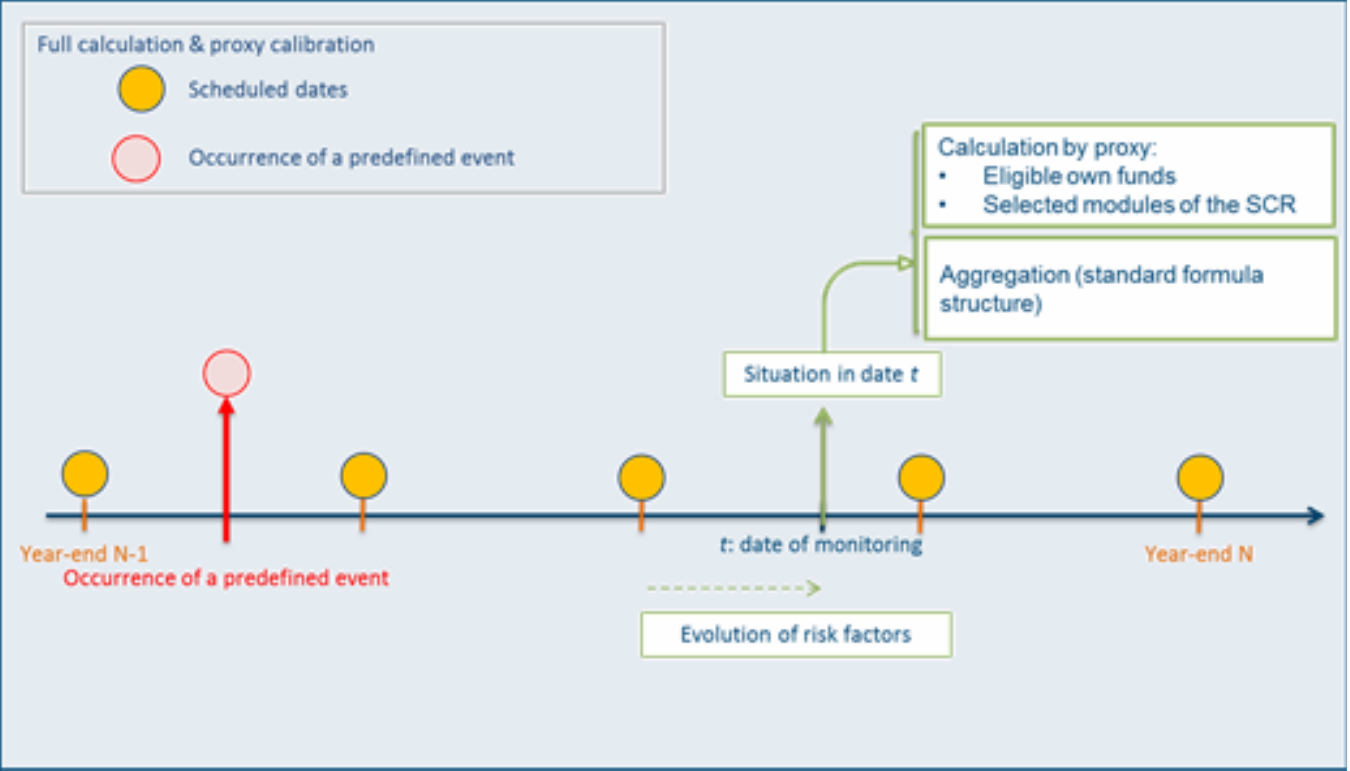}\
\caption{Continuous compliance framework.\label{Figure1}}
\end{figure}

Here below are a few examples of pre-defined events and circumstance,
\begin{itemize}
	\item	external events (\emph{e.g.}: financial events, pandemics),
	\item internal decisions (\emph{e.g.}: change in asset mix),
	\item	risk factors outside the proxies' zone of validity.
\end{itemize}

\section{Quantitative approach to assess the continuous compliance}
\label{QuantitativeApproach}

Note first that the study presented in this paper was carried out in a context where the adjustment for the loss-absorbing capacity of technical provisions was lesser than the Future Discretionary Benefits (``FDB'') (see Level 2 Implementation Measures \cCite{Level22011}). As a consequence, the Value of In-Force and the $NAV$ were always calculated net of the loss-absorbing effect of future profit participation. In cases where the loss-absorbing capacity of technical provisions breaches the FDB, further developments (and additional assumptions), not presented in this paper, will be necessary.

In Section \ref{QuantitativeApproach} we present a proxy implementation that enables one to assess the continuous compliance, and the underlying assumptions.

\subsection{Assumptions underlying the continuous compliance assessment framework}

As explained in Subsection \ref{OverviewApproach}, several simplifications will be necessary in order to operationalize the continuous compliance assessment using our methodology.

\subsubsection{Selection of the monitored risks}

First, we need to assume that the company can be considered subject to a limited number of significant and easily measurable risks with little loss of information.

In most cases this assumption is quite strong. Indeed, there are numerous underlying risks for a life insurance undertaking and these are not always “easily measurable”. For example, the mortality and longevity risks, to cite only those, are greatly difficult to monitor on an infra-year time step, simply because of the lack of data. Moreover the “significant” aspect will have to be justified. For instance, this significance can be defined considering the known impact of the risk on the $SCR$ or on the company's balance sheet, or considering its volatility.

In the case of a life insurance business it seems particularly relevant to select the financial risks, easily measurable and monitorable. As a consequence, the selected risk will for example be the stock, interest rates (corporate, sovereign), implicit volatilities (stock / interest rates), illiquidity premium.

In order to enable a frequent monitoring of the selected risks and of their impact, it is necessary to add the assumption that their evolution over time can be satisfyingly replicated by the evolution of composite indexes defined continuously through the monitoring period.

This assumption is a more tangible translation of the measurable aspect of the risks. The objective here is to enable the risks' monitoring through reference indexes. 

For example, an undertaking which is mainly exposed to European stocks can consider the EUROSTOXX50 level in order to efficiently synthesize its stock level risk. Another possibility may be to consider weighted European stock indexes to obtain an aggregated indicator more accurate and representative of the entity-specific risk. For example, for the sovereign spread risk, it seems relevant for a given entity to monitor an index set up as a weighted average of the spread extracted from the various bonds in its asset portfolio. 

Eventually, the undertaking must aim at developing a indexes table, similar to the following one.
\begin{table}[h!]
\centering
\caption{Example of indexes table --- Significant risks and their associated indicators.}
\begin{tabular}{|l|l|}
	\hline
	 \multicolumn{1}{|c|}{\textbf{Significant risks}} & \multicolumn{1}{c|}{\textbf{Associated composite indicators}} \\
 \hline
	 Stock (level) & 70\% CAC40 / 30\% EUROSTOXX50 \\
	 Risk-free rate (level) & Euro swap curve (averaged level evolution)\\
	 Spread (sovereign) & Weighted average of the spread by issuing country. \\
	 & Weights : \% market value in the asset portfolio\\
	 Spread (corporate) & iTraxx Europe Generic 10Y Corporate\\
	 Volatility (stock) & VCAC Index\\
	 Illiquidity premium & Illiquidity premium (see QIS5 formula \cCite{QIS52010})\\
	\hline
\end{tabular}
\end{table}

Generally speaking, all the assumptions presented here are almost induced by the operational constraints linked to the definition of the continuous compliance framework (full calculation frequency / number of monitored risks). Indeed, it is impossible in practice to monitor each underlying risk day by day. We therefore need to restrict the framework by selecting the most influential risks and indicators enabling their practical monitoring.

In addition, it is irrelevant to consider too stable risks or risks that cannot be monitored infra-annually. In this case, they can simply be assumed frozen, or updated proportionally to a volume measure, through the monitoring period, with little loss of information.

In this simplified framework, a change of the economic conditions over time will be summarize in the realized indexes' level transition. It is then possible to build a proxy enabling one to approximate quickly the $SR$ at each monitoring date, knowing the current level of the composite indicators. 

\begin{figure}[h!]
\centering
\includegraphics[width=0.9\textwidth]{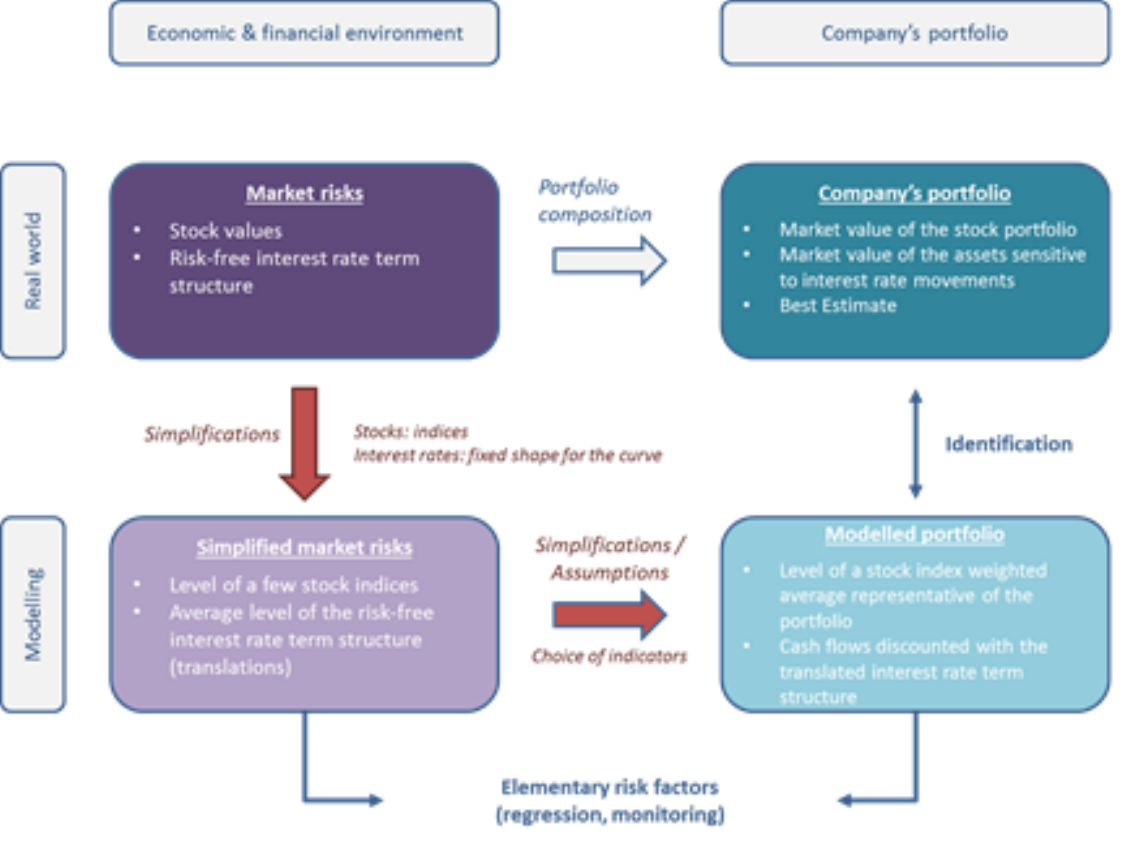}\
\caption{Simplified monitoring framework: Illustration\label{Figure2}}
\end{figure}

Figure \ref{Figure2} illustrates the process to follow and the underlying assumptions made in a simplified framework. Let us develop a case where the company's asset portfolio can be divided into one stock and one bond pools. Two underlying risks have been identified, the stock level risk and the interest rate level risk (average level change of the rates curve\footnote{Note that other kinds of interest rates risks can be selected in order to address the term structure risk more precisely, such as the slope and curvature risks. For more insight on this subject see \emph{e.g.} Diebold and Li \cCite{Diebold2006}.}). Our assumptions lead to consider that, once the risks associated to composite indexes, it is possible to approximate the asset portfolio by a mix between,
\begin{itemize}
	\item	a stock basket with the same returns, composed with the composite stock index only (\emph{e.g.} 70\% CAC 40 / 30\% EUROSTOXX50),
	\item	a bond basket replicating the cash-flows of the bonds discounted using a rate curve induced from the initial curve translated of the average variation of the reference rate curve (the ``composite'' curve, \emph{e.g.} the Euro swap curve).
\end{itemize}

Eventually we can decompose the process presented in Figure \ref{Figure2} between,
\begin{itemize}
	\item a vertical axe where one simplifies the risks themselves,
	\item and an horizontal axe where one transforms the risk into composite indexes.
\end{itemize}

To conclude, note that the assumptions made here will lead to the creation of a basis risk. Indeed, even if the considered indexes are very efficient, one part of the insurance portfolio sensitivity will be omitted due to the approximations. In particular the risks and indexes must be chosen very precisely, entity-specifically. A small mistake can have great repercussions on the approximate $SR$. In order to minimize the basis risk, the undertaking will have to back-test the choices made and the underlying assumptions.

\subsubsection{Selection of the monitored marginal $SCR$}
	
The continuous compliance framework and tool presented in this paper applies to companies that use a Standard Formula approach to assess the $SCR$ value (but can provide relevant information to companies that use an internal model). 

In practice it will not be necessary to monitor every marginal $SCR$ of a company. Indeed, some risk modules will be little or not impacted by any infra-annual evolution of the selected risks. Moreover, a certain number of sub-modules have a small weight in the calculation of the Basic Solvency Capital Requirement ($BSCR$). These too small and/or stable marginal $SCR$ will be frozen or updated proportionally to a volume measure throughout the monitoring period. 

Eventually, the number of risk modules that will have to be updated precisely (the most meaningful marginal $SCR$) should be reduced to less than ten. Note that, among the marginal $SCR$ to recalculate, some can correspond to modeled risks factors but others will not correspond to the selected risk factors while being very impacted by those (\emph{e.g.} the massive lapse $SCR$).

This selection of the relevant $SCR$ sub-modules will introduce a new assumption and a new basis risk, necessary for our methodology's efficiency. The basis risk associated to this assumption, linked to the fact that some marginal $SCR$ will not be updated at each monitoring date, can be reduced by considering a larger number of sub-modules. One will have to apprehend this problem pragmatically, to take a minimal number of risk modules into account in order to limit the number of future calculations, while keeping the error made on the overall $SCR$ under control, the best possible way.

\subsection{Use of parametric proxies to assess the continuous compliance}
\label{ContinuousComplianceProxies}

In the previous section we have defined a reference framework in which we will develop our monitoring tool. The proposed methodology aims at calibrating proxies that replicate the central and shocked $NAV$ as functions of the levels taken by the chosen indexes.

\subsubsection{Assumption of stability of the asset and liability portfolios}

We now work with closed asset and liabilities portfolios, with no trading, claim or premium cash-flow, in order to consider a stable asset-mix and volume of assets and liabilities. Eventually, all the balance sheets movements are now induced by the financial factors.  

This new assumption may seem strong at first sight. However, it seems justified on a short term period. In the general case the evolution of these portfolios is slow for mature life insurance companies. This evolution is therefore assumed to have little significance for the monitoring period of our continuous compliance monitoring tool. Eventually, if a significant evolution happens in practice (\emph{e.g.} a portfolio purchase / sale) this will lead to a full recalibration of the tool (see Subsection \ref{Governance} for more insight on the monitoring tool governance).

\subsubsection{Economic transitions}
	
Let us recall the various assumptions considered until now.
\begin{itemize}
	\item \textsl{H1}: The undertaking's underlying risks can be summarized into a small pool of significant and easily quantifiable risks with little loss of information.
	\item \textsl{H2}: The evolution of these risks can be perfectly replicated by monitoring composite indicators, well defined at each date of the monitoring period.
	\item \textsl{H3}: The number of marginal $SCR$ that need to be precisely updated at each monitoring date can be reduced to the  most impacting risk modules with little loss of information.
	\item \textsl{H4}: The asset and liability portfolio are assumed frozen between two calibration dates of the monitoring tool.
\end{itemize}
	
Under the assumptions \textsl{H1}, \textsl{H2}, \textsl{H3} and \textsl{H4} it is possible to summarize the impact of a time evolution of the economic conditions on the considered portfolio into an instant level shock of the selected composite indicators. This instant choc will be denoted ``economic transition'' and we will see below that it can be identified to a set of elementary risk factors similar to those presented in Devineau and Chauvigny \cCite{Devineau2011}.

\begin{figure}[h!]
\centering
\includegraphics[width=0.8\textwidth]{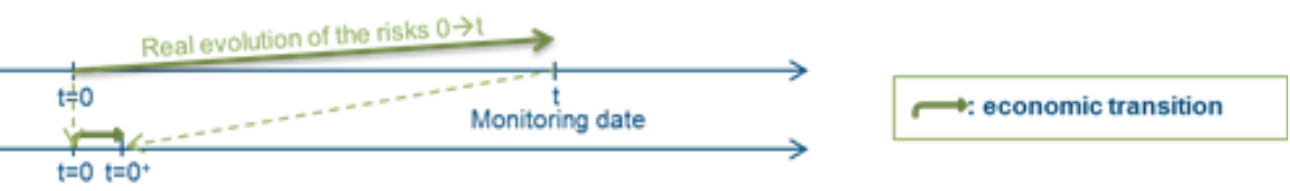}\
\caption{Economic transition ``$0 \rightarrow 0^{+}$''.}
\end{figure}

Let us consider a two shocks framework: the stock level risk, associated to an index denoted by $S(t)$ at date $t\geq0$ ($t=0$ being the tool's calibration date) and an interest rate level risk, associated to zero-coupon prices, denoting by $P(t,m)$ the zero-coupon of maturity $m$ and date $t\geq0$. Now, let us consider an observed evolution between $0$ and a monitoring date $t>0$Finally, to calculate the $NAV$ at date $t$, under our assumptions, it is only necessary to know the new levels $S(t), P(t,m)$.

The real evolution, from $\left(S(0),(P(0,m))_{m\in\llbracket1;M\rrbracket}\right)$ to $\left(S(t),(P(t,m))_{m\in\llbracket1;M\rrbracket}\right)$ can eventually be seen as a risk factors couple,

\begin{equation*}
\varepsilon=\left(^{stock}\varepsilon=\ln \left(\frac{S(t)}{S(0)}\right),^{ZC}\varepsilon=-\frac{1}{M}\sum_{m=0}^{M}\ln \left(\frac{1}{m}\frac{P(t,m)}{P(0,m)}\right)\right),
\end{equation*}

denoting by $^{stock}\varepsilon$ (respectively $^{ZC}\varepsilon$) the stock (resp. zero-coupon) risk factor.

This evolution of the economic conditions, translated into a risk factors tuple, is called economic transition in the following sections of this paper and can easily be extended to a greater number of risks. The risk factor will be used in our algorithm to replicate the instant shocks ``$0 \rightarrow 0^{+}$'' equivalent to the real transitions ``$0 \rightarrow t$''. Moreover, the notion of economic transition will be used to designate either an instant shock or a real evolution of the economic situation between $0$ and $t>0$. In this latter case we will talk about “real” or “realized” economic transition.

\subsubsection{Probable space of economic transitions for a given $\alpha\%$ threshold}

Let us consider, for example, a 3-months monitoring period (with a full calibrations of the monitoring tool at the start and at the end of the period). It is possible to \emph{a priori} assess a probable space of the probable quarterly economic transitions, under the historic probability $\mathbb{P}$ and for a given threshold $\alpha\%$. One simply has to study a deep enough historical data summary of the quarterly evolutions of the indexes and to assess the interval between the $\frac{1-\alpha\%}{2}$ and the $\frac{1+\alpha\%}{2}$ historical quantiles of the risk factors extracted from the historical data set.
	
For example, for the stock risk factor $^stock\varepsilon$, knowing the historical summary $\left(S_{\frac{i}{4}}\right)_{i\in\llbracket0,4T+1\rrbracket}$ one can extract the risk factor's historical data set $\left(^{stock}\varepsilon_{\frac{i}{4}}=\ln \left(\frac{S_{\frac{i+1}{4}}}{S_{\frac{i}{4}}}\right)\right)_{i\in\llbracket0,4T\rrbracket}$ and obtain the probable space of economic transitions for a given $\alpha\%$ threshold, 

\begin{equation*}
\left[q_{\frac{1-\alpha\%}{2}}\left(\left(^{stock}\varepsilon_{\frac{i}{4}}\right)_{i\in\llbracket0;4T\rrbracket}\right);q_{\frac{1+\alpha\%}{2}}\left(\left(^{stock}\varepsilon_{\frac{i}{4}}\right)_{i\in\llbracket0;4T\rrbracket}\right)\right]. 
\end{equation*}
	
In a more general framework, consider economic transitions represented by $J$-tuples of risk factors $\varepsilon=\left(^{1}\varepsilon,...,^{J}\varepsilon\right)$ of which one can get an historical summary $\left(^{1}\varepsilon_{\frac{i}{4}},...,^{J}\varepsilon_{\frac{i}{4}}\right)_{i\in\llbracket0;T\rrbracket}$. The following probable interval of the economic transitions with a $\alpha\%$ threshold can be used,
\begin{center}
$\mathcal{E}^{\alpha}=\left\{\left(^{1}\varepsilon,...,^{J}\varepsilon\right)\in\prod_{j=1}^{J}\left[q_{\frac{1-\alpha\%}{2}}\left(\left(^{j}\varepsilon_{\frac{i}{4}}\right)_{i\in\llbracket0;T\rrbracket}\right);q_{\frac{1+\alpha\%}{2}}\left(\left(^{j}\varepsilon_{\frac{i}{4}}\right)_{i\in\llbracket0;T\rrbracket}\right)\right]\right\}$.
\end{center}

Note that such a space does not take correlations into account. Indeed each risk factor's interval is defined independently from the others. In particular, such a space is \emph{prudent}: contains more than $\alpha\%$ of the historically probable economic evolutions.

\subsubsection{Implementation --- Replication of the central $NAV$}
\label{ReplicationCentralNAV}
	
We will now assume that $J$ different risks have already been selected.

The implementation we will now describe aims at calibrating a polynomial proxy that replicates $NAV_{0^{+}}\left(\varepsilon\right)$, the central $NAV$ at the date $t=0^{+}$, associated to an economic transition $\varepsilon=\left(^{1}\varepsilon,...,^{J}\varepsilon\right)$. The proxy will allow, at each monitoring date $t$, after evaluating the observed economic transition $\varepsilon_{t}$ (realized between $0$ and $t$), to obtain a corresponding approximate central $NAV$ value, $NAV_{0^{+}}^{proxy}\left(\varepsilon_{t}\right)$.
	
\paragraph{Notation and preliminary definitions.} To build the $NAV_{0^{+}}^{proxy}\left(\varepsilon\right)$ function, our approach is inspired from the Curve Fitting ($CF$) and Least Squares Monte Carlo ($LSMC$) polynomial proxies approaches proposed in Vedani and Devineau \cCite{Vedani2013}. It is possible to present a generalized implementation plan for these kinds of approaches. They both aim at approximating the $NAV$ using a polynomial function whose monomials are simple and crossed powers of the elements in $\varepsilon=\left(^{1}\varepsilon,...,^{J}\varepsilon\right)$.
	
Let us introduce the following notation. Let $\mathbf{Q}$ be a risk-neutral measure conditioned by the real-world financial information known at date $0^{+}$, $\mathcal{F}_{0^{+}}$ the filtration that characterizes the real-world economic information contained within an economic transition between dates $0$ and $0^{+}$. Let $R_{u}$ be the profit realized between $u-1$ and $u\geq1$, and $\delta_{u}$ the discount factor at date $u\geq1$ . Let $H$ be the liability run-off horizon.

The gist of the method is described here below. 

The $NAV_{0^{+}}\left(\varepsilon\right)$ depends on the economic information through the period $\left[0;0^{+}\right]$,
\begin{center}
$NAV_{0^{+}}\left(\varepsilon\right)=\mathbb{E}^{\mathbf{Q}}\left[\sum_{t=1}^{H}\delta_{t}R_{t}|\mathcal{F}_{0^{+}}\right]$.
\end{center}

For a given transition $\varepsilon$ it is possible to estimate $NAV_{0^{+}}\left(\varepsilon\right)$ implementing a standard Asset Liability Management model calculation at date $t=0^{+}$. In order to do so one must use an economic scenarios table of $P$ simulations generated under the probability measure $\mathbf{Q}$ between $t=0^{+}$ and $t=H$ initialized by the levels (and volatilities if the risk is chosen) of the various economic drivers as induced by transition $\varepsilon$.
\begin{figure}[h!]
\centering
\includegraphics[width=0.8\textwidth]{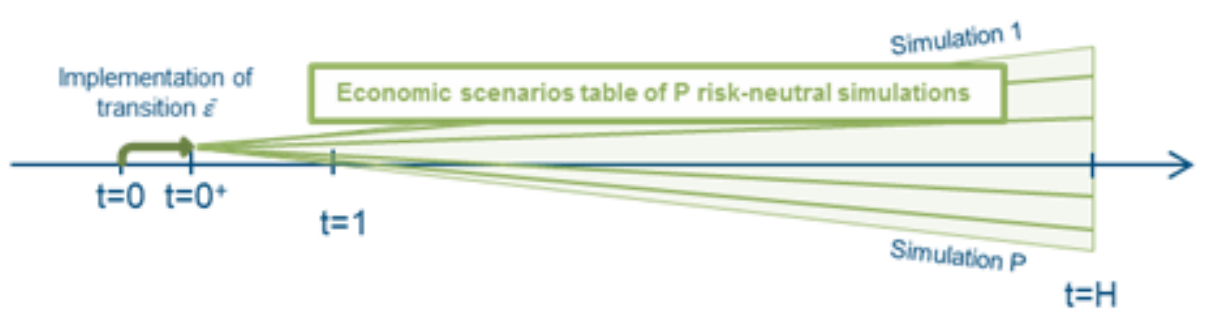}\
\caption{Calculation of an estimator of $NAV_{0^{+}}\left(\varepsilon\right)$ using a Monte Carlo method.}
\end{figure}

For each simulation $p\in\llbracket1;P\rrbracket$ and date $t\in\llbracket1;H\rrbracket$, one has to calculate the profit outcome $R_{t}^{p}$ using an Asset-Liability Management (ALM) model and, knowing the corresponding discount factor $\delta_{t}^{p}$, to assess the Monte Carlo estimator,

\begin{center}
	$\widehat{NAV}_{0^{+}}\left(\varepsilon\right)=\frac{1}{P}\sum_{p=1}^{P}\sum_{t=1}^{H}\delta_{t}^{p}R_{t}^{p}$. 
\end{center}

When $P=1$ we obtain an inefficient estimator of $NAV_{0^{+}}\left(\varepsilon\right)$ which we will denote by $NPV_{0^{+}}\left(\varepsilon\right)$ (Net Present Value of margins), according to the notation of Vedani and Devineau \cCite{Vedani2013}. Note that for a given transition $\varepsilon$, $NPV_{0^{+}}\left(\varepsilon\right)$ is generally very volatile and it is necessary to have $P$ high to get an efficient estimator of $NAV_{0^{+}}\left(\varepsilon\right)$.

\begin{figure}[h!]
\centering
\includegraphics[width=0.7\textwidth]{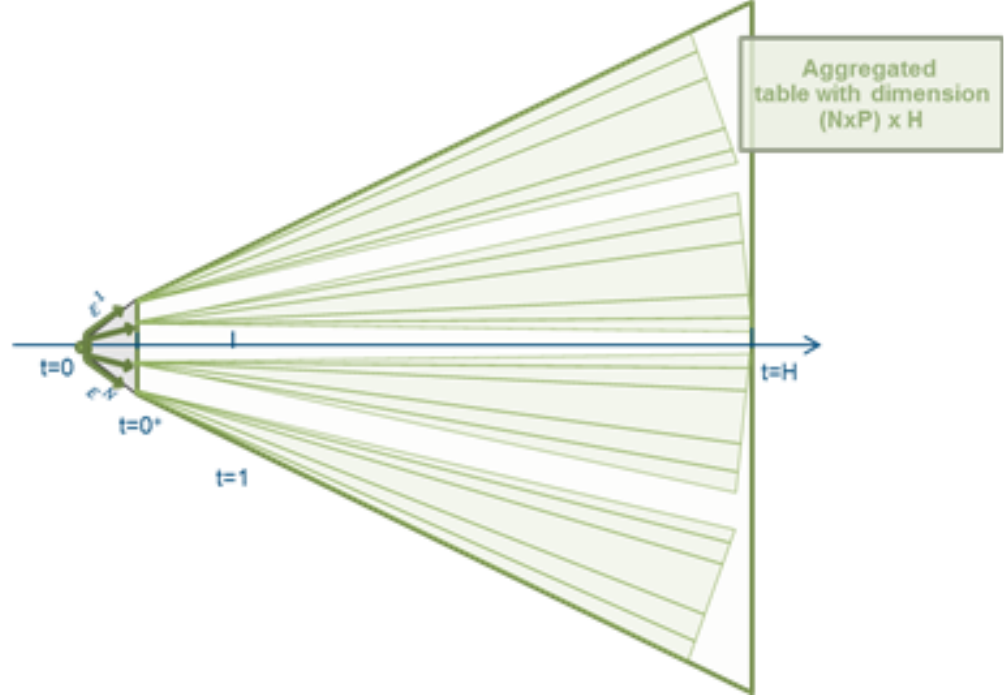}\
\caption{Aggregate table.\label{AggegeteTable}}
\end{figure}

\paragraph{Methodology.} Let us consider a set of $N$ transitions obtained randomly from a probable space of economic transitions and denoted by $\left(\varepsilon^{n}=\left(^{1}\varepsilon^n,...,^{J}\varepsilon^n\right)\right)_{n_\in\llbracket1;N\rrbracket}$. We now have to aggregate all the $N$ associated risk-neutral scenarios tables, each one initialized by the drivers' levels (and volatilities if needed) corresponding to one of the economic transitions in the set, in a unique table (see Figure \ref{AggegeteTable}).

The ALM calculations launched on such a table enables one to get $N \times P$ outcomes
$$\left(NPV_{0^{+}}^{p}\left(\varepsilon^{n}\right)\right)_{n\in\llbracket1;N\rrbracket,p\in\llbracket1;P\rrbracket},$$
and subsequently a $N$ sample
$$\left(NAV_{0^{+}}\left(\varepsilon^{n}\right)=\frac{1}{P}\sum_{p=1}^{P}NPV_{0^{+}}^{p}\left(\varepsilon^{n}\right)\right)_{n\in\llbracket1;N\rrbracket}.$$

Then, the outcomes $\left(\widehat{NAV}_{0^{+}}\left(\varepsilon^{n}\right)\right)_{n\in\llbracket1;N\rrbracket}$ are regressed on simple and crossed monomials of the risk factors in $\varepsilon=\left(^{1}\varepsilon,...,^{J}\varepsilon\right)$. The regression is made by Ordinary Least Squares (OLS) and the optimal regressors $x=\left(\text{Intercept},{}^{1}\!x,...,{}^{K}\!x\right)$ (with, for all $k\in\llbracket1;K\rrbracket$,$^{k}x=\prod_{j=1}^{J}\left(^{j}\varepsilon\right)^{{}^{k}\!\alpha_{j}}$, for a certain $J$-tuple $\left({}^{k}\!\alpha_{1},...,{}^{k}\!\alpha_{J}\right)$ in $\mathbb{N}^{J}$) are selected using a stepwise methodology. For more developments about these approaches see Draper and Smith \cCite{Draper1966} or Hastie \emph{et al.} \cCite{Hastie2009}. 

Let $\beta=\left(^{Int}\beta,{}^{1}\!\beta,...,^{K}\beta\right)$ be the optimal multilinear regression parameters.

The considered multilinear regression can therefore be written under a matricial form $Y=X\beta+U$, denoting by 
\[Y=
\left(\begin{array}{c}
NAV_{0^{+}}\left(\varepsilon^{1}\right)\\
\vdots\\
NAV_{0^{+}}\left(\varepsilon^{N}\right)\\
\end{array}\right),\]
\[X=
\left(\begin{array}{c}
x^{1}\\
\vdots\\
x^{N}\\
\end{array}\right),\]
with, for all $n\in\llbracket1;N\rrbracket$, $x^{n}=\left(1,{}^{1}\!x^{n},...,^{K}x^{n}\right)$, for all $k\in\left(\left|1;K\right|\right)$, $^{k}x^{n}=\prod_{j=1}^{J}\left(^{j}\varepsilon^{n}\right)^{{}^{k}\!\alpha_{j}}$ and $U=Y-X\beta$.
	
In this regression, the conditional expectation of $NAV_{0^{+}}\left(\varepsilon^{n}\right)$ given the $\sigma$-field generated by the regressors matrix $X$ is simply seen as a linear combination of the regressors. For more insight about multiple regression models the reader may consult Saporta \cCite{Saporta2006}. 

The underlying assumption of this model can also be written $\exists\beta,\mathbb{E}\left[Y|X\right]=X\beta$. 

Under the assumption that $X'X$ is invertible (with $Z'$ the transposition of a given vector or matrix $Z$), the estimated vector of the parameters is,\
\begin{center}
$\hat{\beta}=\left(X'X\right)^{-1}X'Y.$
\end{center}

Moreover, for a given economic transition $\bar{\varepsilon}$ and its associated set of optimal regressors $\bar{x}$, $\bar{x}.\hat{\beta}$ is an unbiased and consistent estimator of $\mathbb{E}\left[\widehat{NAV}_{0^{+}}\left(\bar{\varepsilon}\right)|\bar{x}\right]=\mathbb{E}\left[NAV_{0^{+}}\left(\bar{\varepsilon}\right)|\bar{x}\right]$. When $\sigma\left(x\right)=\mathcal{F}_{0^{+}}$, which is generally the case in practice, $\bar{x}\hat{\beta}$ is an efficient estimator of $NAV_{0^{+}}\left(\varepsilon\right)$ and we get an efficient polynomial proxy of the central $NAV$ for every economic transition. 

Eventually, it is necessary to test the goodness of fit. The idea is now to calculate several approximate outcomes of central $NAV$, associated to an \emph{out of sample}\footnote{Scenarios that are not included in the set used during the calibration steps.}  set of economic transition, using a Monte Carlo method on a great number of secondary scenarios, and to compare these outcomes to those obtained using the proxy.   
	
\subsubsection{Implementation --- Replication of the shocked $NAV$}

At each monitoring date, we aim at knowing each pertinent marginal $SCR$ value, for each chosen risk modules. With the proxy calibrated in the previous section one can calculate an approximate value of the central $NAV$. We now have to duplicate the methodology presented in Subsection \ref{ReplicationCentralNAV}, adapted for each marginally shocked $NAV$ (considering the Standard Formula shocks).\footnote{Note that it is necessary to calibrate new ``after shock'' proxies because it is impossible to assimilate a Standard Formula shock to a transition shock...}
	
The implementation is fully similar except the fact that the shocked proxies are calibrated on $N$ outcomes of marginally shocked $NAV_{0^{+}}$. Indeed each marginal $SCR$ is a difference between the central $NAV$ and a $NAV$ after the application of the marginal shock. We therefore need the $NAV$ after shock that takes the conditions associated to an economic transition into account.

This enables one to obtain, for each shock $nb~i$, a set 
$\left(NAV_{0^{+}}^{shock~nb~i}\left(\varepsilon^{n}\right)\right)_{n\in\llbracket1;N\rrbracket}$, a new optimal regressors set $\left(\text{Intercept}, {}^{1}\!x^{shock~nb~i},...,{}^{K}\!x^{shock~nb~i}\right)$ and new optimal parameters estimators $\hat{\beta}^{shock~nb~i}$.

\subsubsection{Practical monitoring}
\label{PracticalMonitoring}

Once the methodology has been implemented, the obtained polynomial proxies enable one, at each date within the monitoring period, to evaluate the central and shocked $NAV$ values knowing the realized economic transition.
	
At each monitoring date $t$, the process is the following.
\begin{itemize}
	\item Assessment of the realized transition between $0$ and $t$, $\hat{\varepsilon}$.
	\item Derivation of the values of the optimal regressors set for each proxy:
	\begin{itemize}
		\item $\bar{x}$ the realized regressors set for the central proxy,
		\item $\bar{x}^{shock~nb~1},...,\bar{x}^{shock~nb~J}$ the regressors set for the $J$ shocked proxy.
	\end{itemize}
	\item Calculation of the approximate central and shocked $NAV$ levels at date $t$:
	\begin{itemize}
		\item $\bar{x}\hat{\beta}$, the approximate central $NAV$,
		\item $\bar{x}^{shock~nb~1}\hat{\beta}^{shock~nb~1},...,\bar{x}^{shock~nb~J}\hat{\beta}^{shock~nb~J}$ the $J$ approximate shocked $NAV$.
	\end{itemize}
	\item Calculation of the approximate marginal $SCR$ and, considering frozen values, or values that are updated proportionally to a volume measure, for the other marginal $SCR$, Standard Formula aggregation to evaluate the approximate overall $SCR$ and $SR$\footnote{For more insight concerning the Standard Formula aggregation, especially about the evaluation of the differed taxes, see Subsection \ref{StandardFormulaAggregationProcess}.}.
\end{itemize}

\subsection{Least-Squares Monte Carlo vs. Curve Fitting --- The large dimensioning issue}
	
The implementation developed in Subsection \ref{ContinuousComplianceProxies} is an adapted application, generalized to the $N \times P$ framework, of the polynomial approaches such as $LSMC$ and $CF$, already used in previous studies to project $NAV$ values at $t$ years ($t\geq1$). For more insight about these approaches, see for example Vedani and Devineau \cCite{Vedani2013}, Algorithmics \cCite{Algorithmics2011} or Barrie \& Hibbert \cCite{Barrie2011}.

When $P=1$ and $N$ is very large (basically the proxies are calibrated on Net Present Values of margins / $NPV$), we are in the case of a $LSMC$ approach. On the contrary, when $N$ is rather small and $P$ large, we are in the case of a $CF$ approach.

Both approaches generally deliver similar results. However the $LSMC$ is often seen as more stable than a $CF$ when a large number of regressors are embedded in the proxy. This clearly matches the continuous compliance case where the user generally considers a larger number of risk factors compared to the usual $LSMC$ methodologies, used to accelerate Nested Simulations for example. In our case, this large dimensioning issue makes a lot of sense.

In Section \ref{LSMC_ContinuousCompliance} we will apply the methodology on four distinct risk factors, the stock level risk, the interest rates level risk, the widening of corporates spread and of sovereign spread risks. We have chosen to implement this application using a $LSMC$ method. In Section $5$ we eventually try to challenge the commonly agreed idea that this methodology is more robust than $CF$ in a large dimension context.

\section{$LSMC$ approach adapted to the continuous compliance issue}
\label{LSMC_ContinuousCompliance}
	
In Section \ref{LSMC_ContinuousCompliance} we will implement the presented methodology, in a standard savings product framework. The ALM model used for the projections takes profit sharing mechanisms, target crediting rate and dynamic lapses behaviors of policy holders into account. Its characteristics are similar to those of the model used in Section 5 of Vedani and Devineau \cCite{Vedani2013}. The economic assumptions are those of 31/12/2012.

\subsection{Implementation of the monitoring tool --- Initialization step and proxies calibration}

Firstly it is necessary to shape the exact framework of the study. We have to select the significant risks to be monitored, to choose representative indexes and then to identify the risk modules that will be updated. Note that the other risk modules will be considered frozen through the monitoring period.

The monitoring period must be chosen short enough to ensure a good validity of our stability assumptions for the risk modules that are not updated and for the balance sheet composition. However, it also defines the time during two complete proxy calibrations and, as a consequence, it must be chosen long enough not to force too frequent calibrations, which are highly time-consuming. In this study we have therefore chosen to consider a quarterly monitoring period.

\subsubsection{Initialization step --- Implementation of a complete regulatory solvency calculation}

In order to quantify the relative relevance of the various marginal $SCR$ of the Standard Formula, it is recommended to implement, as a preliminary step, a complete regulatory solvency calculation before a calibration of the monitoring tool. Moreover, seen as an \emph{out of sample} scenario, this central calculation can be used as a validation point for the calibrated proxies.\footnote{The implementation of two to four complete regulatory solvency calculations may be a strong constraint for most insurance undertakings however, due to the several assumptions made to implement the monitoring tool, we recommend to consider monitoring period no longer than six months.}

It is also possible to select the marginal $SCR$ based on expert statements or on the undertaking's expertise, knowing the products sensitivities to the various shocks and economic cycles at the calibration date (and the previous $SCR$ calculations).

\subsubsection{Initialization step --- Risk factor and monitored indexes selection}

We have selected four major risks and built the following indexes table.
\begin{table}[h!]
\centering
\caption{Selected risks and associated indicators.}
\begin{tabular}{|l|l|}
	\hline
	 \multicolumn{1}{|c|}{\textbf{Selected risks}} & \multicolumn{1}{c|}{\textbf{Composite indicators}} \\
 \hline
	 Stock (level) & 100\% EUROSTOXX50 \\
	 Risk-free rate (level) & Euro swap curve (averaged level evolution)\\
	 Spread (sovereign) & Average spread French bonds rate vs. Euro swap rate\\
	 Spread (corporate) & iTraxx Europe Generic 10Y Corporate\\
	\hline
\end{tabular}
\end{table}

These four risks generally have a great impact on the $NAV$ and $SCR$ in the case of savings products, even on a short monitoring period. Moreover, they are highly volatile at the calibration date (31/12/12). In particular, the division of the spread risk in two categories (sovereign and corporate) is absolutely necessary within the European sovereign debt context.

A wide range of risks have been set aside of this study that is just intended to be a simple example. In practice both the stock and interest rates implicit volatility risks are also relevant risks that can be added in the methodology's implementation with no major issue. For the stock implicit volatility risk it is possible to monitor market volatility indexes such as the VIX. Note that the interest rates implicit volatility risk raises several questions related to the application of the risk in the instant economic transitions, in the calibration scenarios. These issues can be set aside considering recalibration/regeneration approaches (see Devineau \cCite{Devineau2010}) and will not be discussed in this paper.

\subsubsection{Initialization step --- Choice of the monitored marginal $SCR$}

Considering the updated risk modules to update, we have chosen the most significant in the Standard Formula aggregation process (see Table \ref{TableSCR}). These are also the less stable trough time, 
\begin{itemize}
	\item the stock $SCR$,
	\item	the interest rates $SCR$,
	\item	the spread $SCR$,
	\item	the liquidity $SCR$.
\end{itemize}

The lapse risk $SCR$, generally highly significant, has not been considered here. Indeed with the very low rates, as at 31/12/2012, the lapse risk $SCR$ is close to zero. Some other significant $SCR$ sub-modules such as the real estate $SCR$ have been omitted because of their low infra-year volatility. 
\clearpage
\begin{table}[h!]
\centering
\caption{Market marginal $SCR$ as at 31/12/2012.\label{TableSCR}}
\begin{tabular}{| l | r c|}
	\hline
	 \multicolumn{1}{|c|}{\textbf{Market $SCR$}} & \multicolumn{2}{c|}{\textbf{Value as at 31/12/2012}} \\
 \hline
	 IR $SCR$ & \hspace*{12ex} 968 & \\
	 Stock $SCR$ & 3930 & \\
	 Real Estate $SCR$ & 943 & \\
	 Spread $SCR$ & 2658 & \\
	 Liquidity $SCR$ & 3928 & \\
	 Concentration $SCR$ & 661 & \\
	 Currency $SCR$ & 127 & \\
	 \multicolumn{1}{|c|}{...} & ... & \\
	\hline
\end{tabular}
\end{table}

\subsubsection{Proxies calibration and validation}

The calibration of the various proxies is made through the same process as developed in Vedani and Devineau \cCite{Vedani2013}. The proxy is obtained by implementing a standard $OLS$ methodology and the optimal regressors are selected through a stepwise approach. This enables the process to be completely automated. The validation of each proxy is made by considering ten \emph{out of the sample} scenarios. These are scenarios that have not be used to calibrate the proxies but on which we have calculated shocked and central outcomes of $\widehat{NAV}_{0^{+}}$. These ``true'' outcomes are then compared to the approximate outcomes obtained from our proxies.

To select the \emph{out of the sample} scenarios we have chosen to define them as the 10 scenarios that go step by step from the ``initial'' position to the ``worst case'' situation (the calibrated \emph{worst case} limit of the monitored risks).

For each risk factor $\epsilon^{j}$,
\begin{itemize}
	\item the ``initial'' position is $\epsilon^{j}_{init}=0$,
	\item the ``worst case'' situation\footnote{= the $10^{th}$ \emph{out of sample} scenario} is $\epsilon^{j}_{w.c.}=q_{\frac{1-\alpha\%}{2}}\left(\left(\varepsilon_{\frac{i}{4}}\right)_{i\in\llbracket0;T\rrbracket}\right)$ or $q_{\frac{1+\alpha\%}{2}}\left(\left(\varepsilon_{\frac{i}{4}}\right)_{i\in\llbracket0;T\rrbracket}\right)$, depending on the direction of the worst case for each risk,
	\item the $k^{th}$ ($k\in\llbracket1;9\rrbracket$) \emph{out of sample} scenario is $\epsilon^{j}_{nb.~k}=\frac{k}{10}\epsilon^{j}_{w.c.}+\frac{10-k}{10}\epsilon^{j}_{init}$.
\end{itemize}

Below (in Table \ref{Valid}) are shown the relative deviations, between the proxies outcomes and the corresponding \emph{out of sample} fully-calculated scenarios, obtained on the first five validation scenarios. 

As one can see, the relative deviations are always close to $0$ apart from the illiquidity shocked $NAV$ proxy. In practice this proxy is the most complex to calibrate due to the high volatility of the illiquidity shocked $NAV$. To avoid this issue, the user can add more calibration scenarios or select more potential regressors when implementing the stepwise methodology. In our study we have chosen to validate our proxy, staying critical on the underlying approximate marginal $SCR$ illiquidity.  

We do not discuss in this paper the optimal way to select the validation scenarios. More generally we acknowledge the need for a reflection about the calculation error due to the use of proxies in Life insurance projections. This reflection is underway and will be the subject of future articles. In the framework presented in this article, the analysis of the proxies’ accuracy could be threefold:
\begin{itemize}
\item validation of the polynomials on the basis of out-of-sample scenarios (as mentioned above, the scenario selection process will be investigated further in the future);
\item at different dates during the monitoring period, update the proxies and run simplified full calculations on the basis of the risk factors’ observed values. By doing this, we are adding new out-of-sample scenarios, selected ex-post for their relevance. Note that in the simplified full calculations, the only inputs updated are the risk factors selected for the proxies;
\item comparison at different dates (prior to and after the calibration dates) between the results obtained with the proxies and those obtained by full calculations, with all the inputs of the full calculations being updated. This would be a way to confirm that the risk factor selection is satisfactory. For example, if the proxies are calibrated at dates $t_{-1}$, $t_{0}$ and $t_{1}$, the results obtained with the proxies calibrated in $t_{0}$ could be compared with the full calculations at dates $t_{-1}$, $t_{0}$ and $t_{1}$.
\end{itemize}
Note however that the number of tests performed will have to be limited so that the implementation of proxies remains operationally beneficial.

\begin{table}[h!]
\centering
\caption{Relative deviations proxies vs. full-calculation $NAV$ (check on the five first validation steps).\label{Valid}}
\begin{tabular}{|l|r|r|r|r|r|}
	\hline
	 \multicolumn{1}{|c|}{\textbf{Validation scenarios}} & \multicolumn{1}{c|}{\textbf{1}} & \multicolumn{1}{c|}{\textbf{2}} & \multicolumn{1}{c|}{\textbf{3}} & \multicolumn{1}{c|}{\textbf{4}} & \multicolumn{1}{c|}{\textbf{5}} \\
 \hline
	 Central $NAV$ & -0.07\% & 1.65\% & 1.56\% & 1.05\% & 0.29\%\\
	 IR shocked $NAV$ & -0.18\% & 1.67\% & 1.14\% & 0.44\% & -0.83\%\\
	 ``Global'' Stock shocked $NAV$ & 0.24\% & 1.93\% & 1.56\% & 1.15\% & 0.28\%\\
	 ``Other'' Stock shocked $NAV$ & 0.19\% & 1.95\% & 1.78\% & 1.31\% & 0.27\%\\
	 Spread shocked $NAV$ & 0.01\% & 2.29\% & 2.15\% & 1.06\% & 0.18\%\\
	 Illiquidity shocked $NAV$ & -5.35\% & -3.27\% & -2.43\% & -3.03\% & -2.39\%\\
	\hline
\end{tabular}
\end{table}

All the proxies being eventually calibrated and validated, it is now necessary to rebuild the Standard Formula aggregation process in order to assess the approximate overall $SCR$ value.

\subsubsection{Proxies aggregation through the Standard Formula process}
\label{StandardFormulaAggregationProcess}

In practice the overall $SCR$ is calculated as an aggregation of three quantities, the $BSCR$, the operational $SCR$ ($SCRop$) and the tax adjustments ($Adj$).

As far as the $BSCR$ is concerned, no particular issue is raised by its calculation. At each monitoring date, the selected marginal $SCR$ are approximated using the proxies and the other $SCR$ are assumed frozen. The $BSCR$ is simply obtained through a Standard Formula aggregation (see for example Devineau and Loisel \cCite{Devineau2009b}).

To derive the operational $SCR$, we consider that this capital is also stable through time, which is in practice an acceptable assumption for a half-yearly or quarterly monitoring period (and consistent with the asset and liability portfolios stability assumption).

The Tax adjustments approximation leads to the greatest issue. Indeed we need to know the approximate Value of In-Force ($VIF$) at the monitoring date. We obtain the approximate $VIF$ as the approximate central $NAV$ $\left(\widehat{NAV}^{central proxy}\right)$ minus a fixed amount calculated as the sum of the tier-one own funds ($tier\_one\_OF$) and of the subordinated debt ($SD$) minus the financial management fees ($FMF$), as at the calibration date. Let $t$ be the monitoring date and $0$ be the proxies' calibration date ($t>0$),
\begin{center}
$\widehat{VIF}_{t}\approx NAV_{t}^{central~proxy}-\left(tier\_one\_OF_{0}+SD_{0}-FMF_{0}\right)$.
\end{center}

Assuming a frozen corporation tax rate of $34.43\%$ (French corporation tax rate), the approximated level of deferred tax liability $\widehat{DTL}$ is obtained as,
\begin{center}
$\widehat{DTL}_{t}=34.43\% \times \widehat{VIF}_{t}$.
\end{center}

Eventually, the income tax recovery associated to new business $\left(ITR^{NB}\right)$ is assumed frozen through the monitoring period and the approximate tax adjustments at the monitoring date is obtained as,
\begin{center}
$\widehat{Adj}_{t}=ITR^{NB}_{0}+\widehat{DTL}_{t}$.
\end{center}

Knowing the approximate values $\widehat{BSCR}_{t}$ and $\widehat{Adj}_{t}$, and the initial value $SCRop_{0}$, one can obtain the approximate overall $SCR$ (simply denoted by $\widehat{SCR}$) at the monitoring date as,
\begin{center}
$\widehat{SCR}_{t}=\widehat{BSCR}_{t}+SCRop_{0}-\widehat{Adj}_{t}$.
\end{center}

Eventually, in order to obtain the $SR$ approximation we obtain the approximate eligible own funds $\widehat{OF}$ as,
\begin{center}
$\widehat{OF}_{t}=\left(tier\_one\_OF_{0}+SD_{0}-FMF_{0}\right)+\widehat{VIF}_{t} \times \left(1-34.43\%\right)$.
\end{center}

Eventually, the approximate $SR$ at the monitoring date is,
\begin{center}
$\widehat{SR}_{t}=\frac{\widehat{OF}_{t}}{\widehat{SCR}_{t}}$.
\end{center}

\subsection{Practical use of the monitoring tool}
\label{PracticalUse}

In subsection \ref{PracticalUse} we will first see the issues raised by the practical continuous compliance's monitoring through our tool, and the tool's governance. In a second part we will develop the other possible uses of the monitoring tool, especially in the area of the risk management and for the development of preventive measures. 

\subsubsection{Monitoring the continuous compliance}

At each monitoring date the process to assess the regulatory compliance is the same as presented in Subsection \ref{PracticalMonitoring}.
\begin{itemize}
	\item	Assessment of the realized transition between $0$ and $t$, $\hat{\varepsilon}$.
	\item	Derivation of the values of the optimal regressors set for each proxy.
	\item	Derivation of the values of the optimal regressors set for each proxy.
	\item Calculation of the approximate central and shocked $NAV$ levels at date $t$.
	\item	Calculation of the levels of each approximate marginal $SCR$ at date $t$ (the other marginal $SCR$ are assumed frozen through the monitoring period). 
\end{itemize}

This, with other stability assumptions such as stability of the tax rate and of the tier-one own funds, enables one to reconstruct the Basic $SCR$, the operational $SCR$ and the Tax adjustments and, eventually, to approximate the overall $SCR$ and the $SR$ at the monitoring date.

\begin{figure}[h!]
\centering
\includegraphics[width=0.75\textwidth]{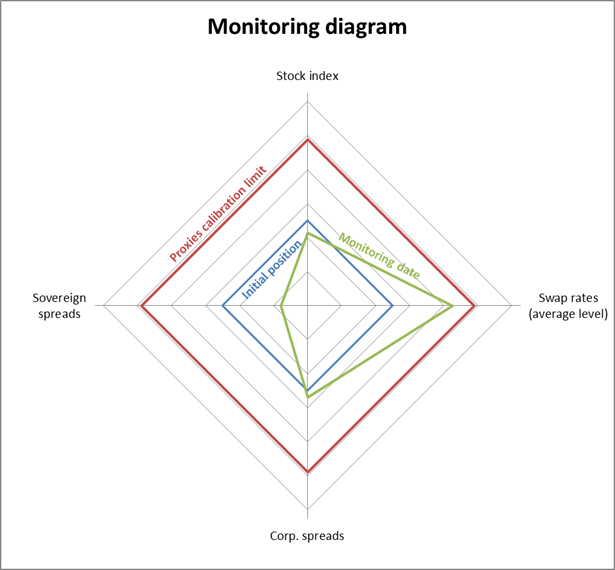}
\caption{Diagram used to monitor the evolution of the risk factors \label{Target}}
\end{figure}

Note that this process can be automated to provide a monitoring diagram such as the one depicted below and a set of outputs such as the eligible own funds, the overall $SCR$, the $SR$, but also the various marginal $SCR$ (see Figure \ref{Target}).

\subsubsection{Monitoring the daily evolution of the $SR$}

\begin{figure}[h!]
\centering
\includegraphics[width=0.95\textwidth]{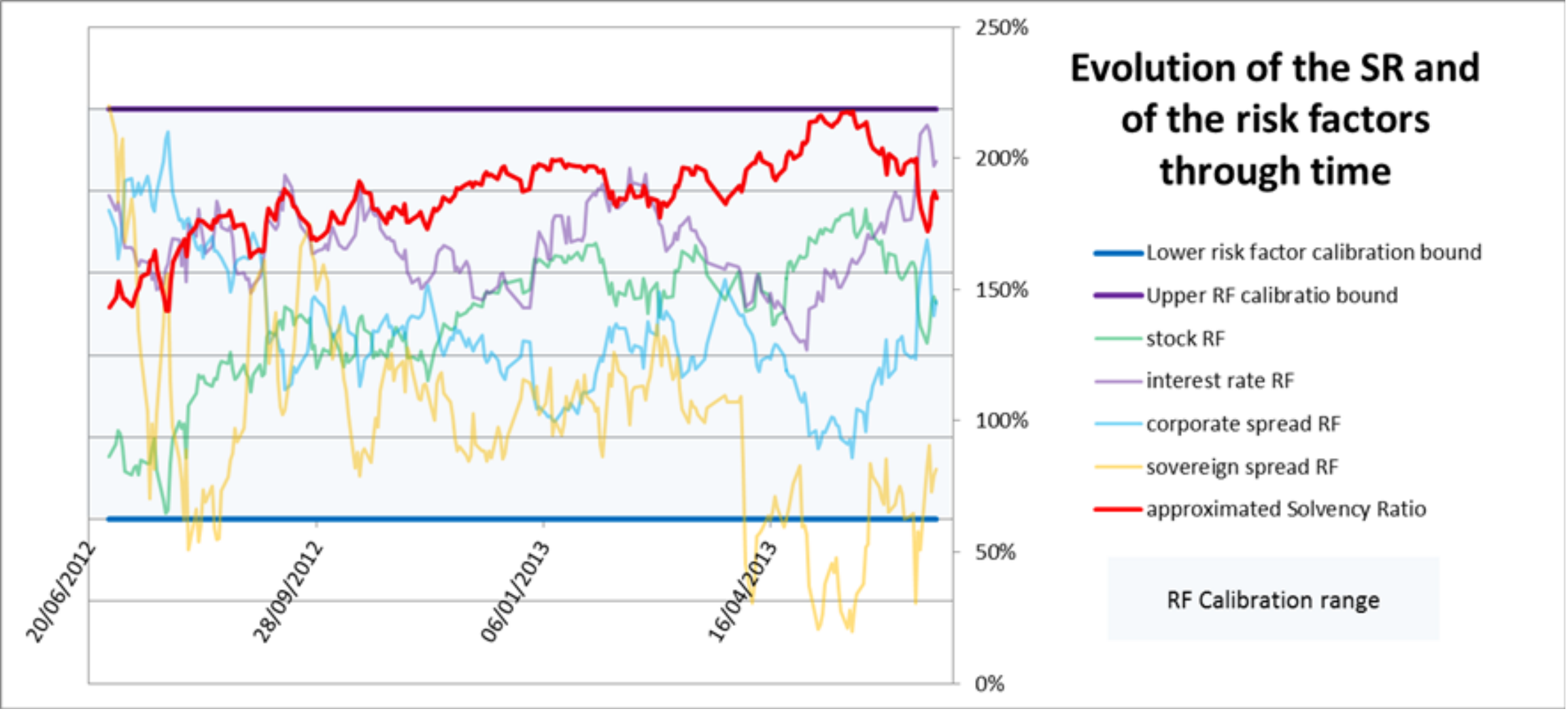}
\caption{Monitoring of the approximate $SR$ and of the four underlying risk factors, from 30/06/12 to 30/06/13.\label{Fig7}}
\end{figure}

In practice the ability to monitor the $SR$ day by day is very interesting and provides a good idea of the empirical volatility of such a ratio (see Figure \ref{Fig7}).

\begin{figure}[h!]
\centering
\includegraphics[width=0.95\textwidth]{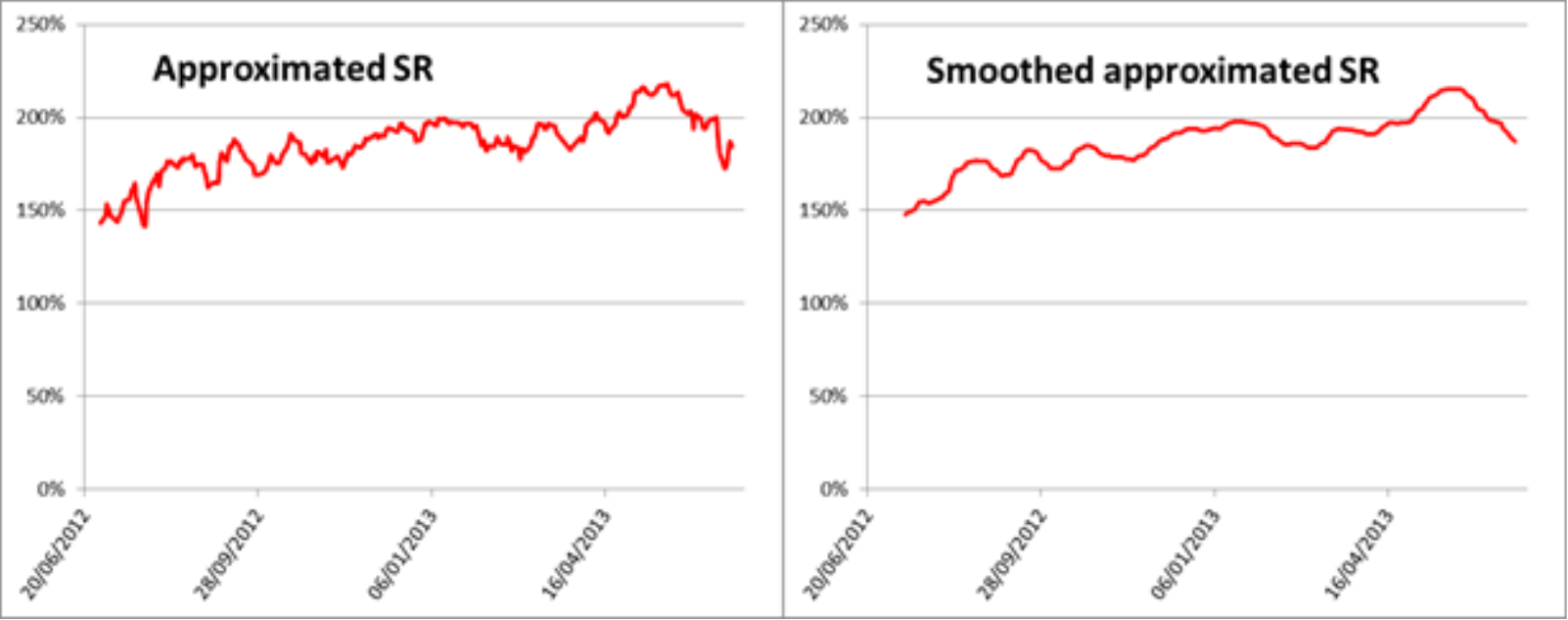}
\caption{Comparison of the standard approximate $SR$ and of a smoothed approximate $SR$ - Monitoring from 30/06/12 to 30/06/13.\label{Fig8}}
\end{figure}

In particular, in an ORSA framework it could be relevant to consider an artificially smoothed $SR$, for example using a 2-week moving average, in order to depict a more consistent solvency indicator. Considering the same data as presented in the previous figure we would obtain the following two graphs (see Figure \ref{Fig8}).

\subsubsection{Monitoring tool governance}
\label{Governance}

Several assumptions are made to provide the approximate $SR$ but we can observe in practice a good replication of the risk impacts and of the $SR$ variations. However the use of this monitoring tool only provides a proxy and therefore the results must be used with caution and its governance must be managed very carefully.
\\

The governance of the tool can be divided into three parts. 
\begin{itemize}
	\item	Firstly it is necessary to \emph{a priori} define the recalibration frequency. The monitoring period associated to each total calibration of the tool should not be too long. The authors believe it should not exceed half a year.
	\item Secondly it is important to identify clearly the data to update for each recalibration. These data especially cover the asset and liability data.
	\item Finally the user must define the conditions leading to a total (unplanned) recalibration of the tool. In particular, these conditions must include updates following management decisions (financial strategy changes inside the mode, asset mix changes,...) and updates triggered by the evolution of the economic situation.
\end{itemize}

\subsubsection{Alternative uses of the tool}

This monitoring tool enables the risk managers to run a certain number of studies, even at the beginning of the monitoring period, in order ton anticipate the impact of future risk deviations for example.

\begin{figure}[h!]
\centering
\includegraphics[width=1\textwidth]{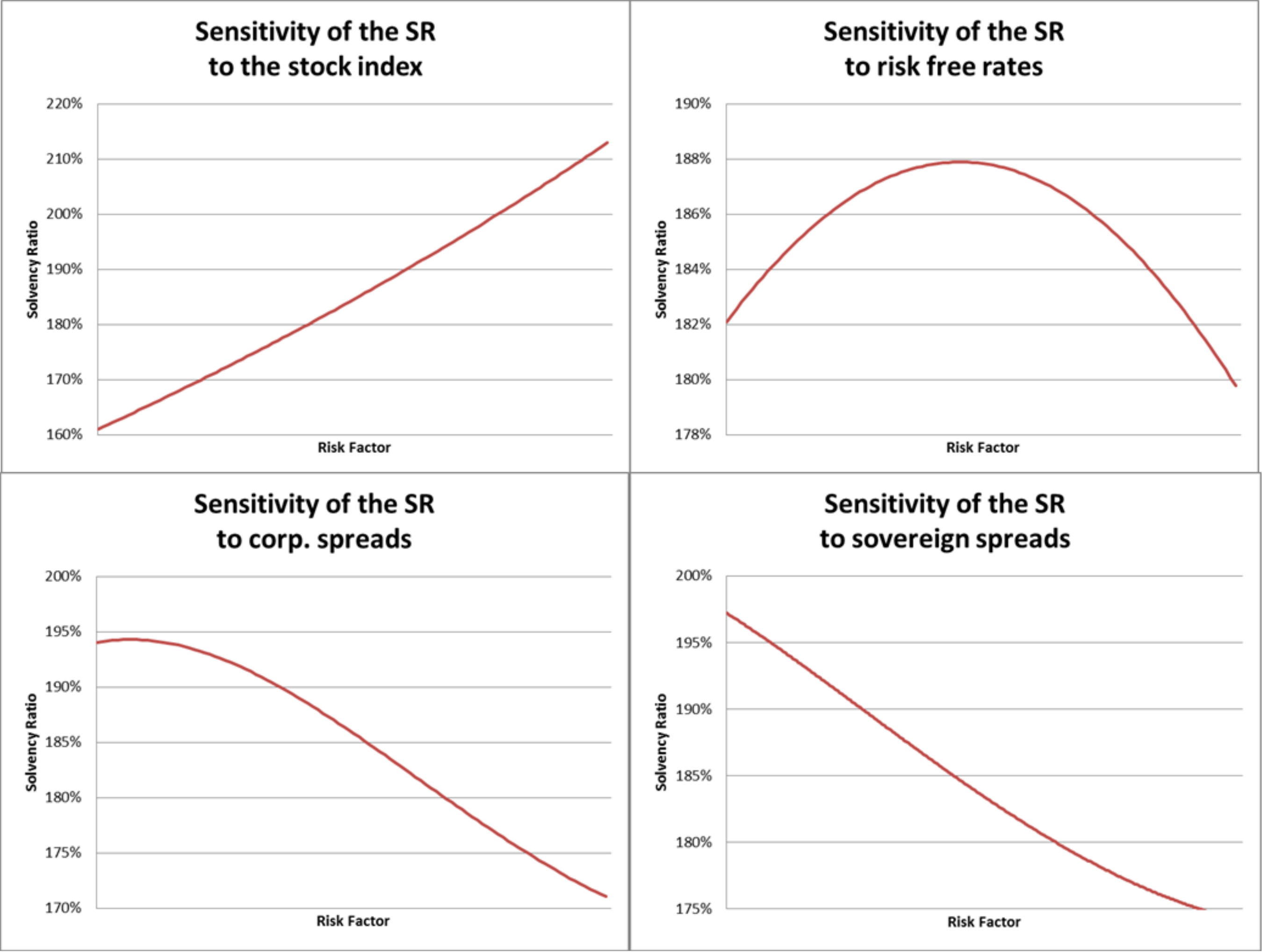}
\caption{1D solvency ratio sensitivities.\label{Fig9}}

\vspace{5ex}

\includegraphics[width=1\textwidth]{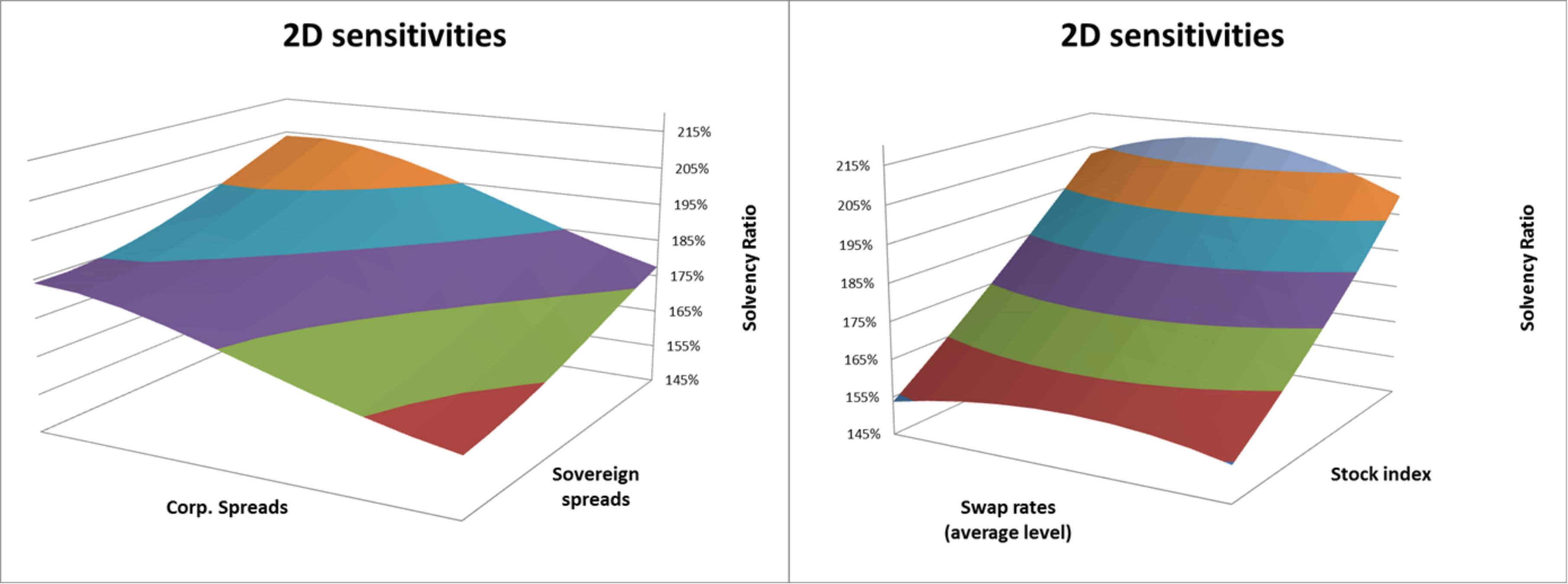}
\caption{2D solvency ratio sensitivities.\label{Fig10}}
\end{figure}
\clearpage

\paragraph{Sensitivity study and stress testing.}

The parametric proxy that replicates the central $NAV$ can also be used to stress the marginal and joint sensitivities of the $NAV$ to the various risks embedded in our proxies. Even more interesting for the risk managers, it is possible to assess a complete sensitivity study directly on the $SR$ of the company, which is very difficult to compute without using an approximation tool (see Figures \ref{Fig9} and \ref{Fig10}).

This sensitivity analyzes needs no additional calculations to the proxies' assessment and enables the risk managers to compute as many ``approximate'' stress tests as needed. In practice such a use of the tool enables to gain better insight about the impact of each risk, taken either individually or jointly, on the $SR$.

\begin{figure}[h!]
\centering
\includegraphics[width=0.8\textwidth]{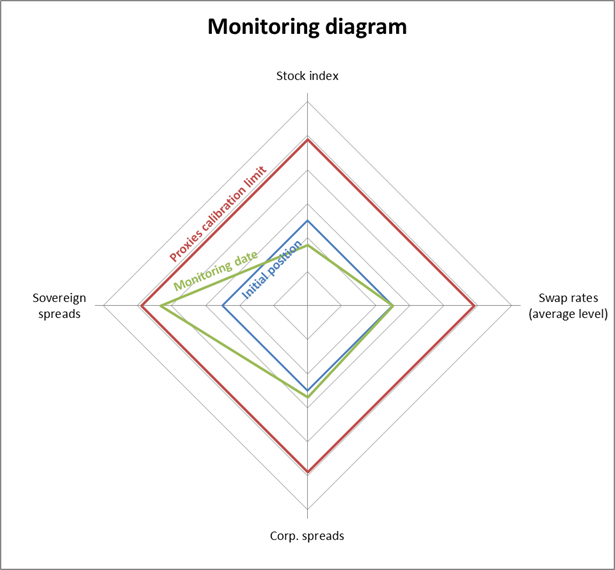}
\caption{Monitoring diagram after the fictitious evolution of the monitored risks.\label{Fig11}}
\end{figure}

\begin{figure}[h!]
\centering
\includegraphics[width=1\textwidth]{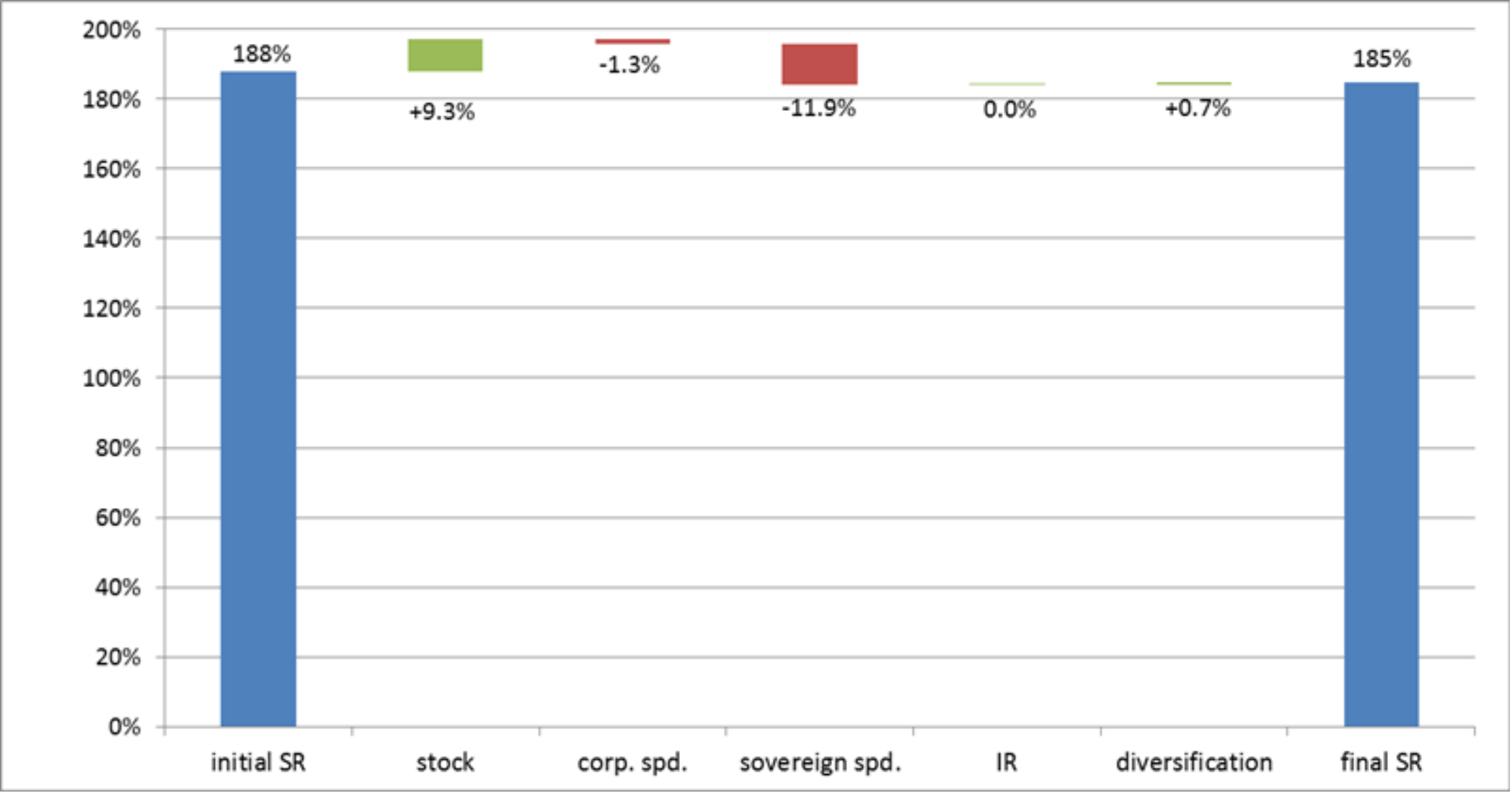}
\caption{Marginal impact of the risks on the $SR$ between the calibration date and a “virtual” monitoring date.\label{Fig12}}
\end{figure}

\paragraph{Monitoring the marginal impacts of the risks and market anticipations.} Using our monitoring tool it is possible to trace the evolution of the $SR$ risk after risk (only for the monitored risks). Figures \ref{Fig11} and \ref{Fig12} correspond to a ficticious evolution of the risks implemented between the calibration date and a ``virtual'' monitoring date). 

Such a study can be run at each monitoring date, or on fictitious scenarios (\emph{e.g.} market anticipations), in order to provide better insight about the $SR$ movements through time.

Concerning market anticipations, if a risk manager anticipates a rise or a fall of the stocks / interest rates / spread, he can directly, through our tool, evaluate the corresponding impact on the undertaking’s $SR$. In particular, such a study can be relevant to propose quantitative preventive measures.

In practice it also seems possible for the user to add asset-mix weights in the monitored risk factors set. This would enable the user to \emph{a priori} test asset-mix re-balancing possibilities in order to select specific preventive measures and prevent the worse market anticipation. This implementation has not been done yet but will be part of the major future developments of the monitoring tool.

\section{Empirical comparison between $LSMC$ and $CF$}
\label{LSMCandCF_Comparison}

In Section \ref{LSMCandCF_Comparison} we will try to challenge the generally agreed idea that the $LSMC$ methodology is more robust than $CF$ in a large dimension context. We will also consider the various possibilities to build asymptotic confidence intervals for the obtained $NAV$ estimators using our polynomial proxies.

\subsection{Mathematical formalism --- Probabilistic vs. statistic framework}
\label{MathematicalFormalization}

Through Subsection \ref{MathematicalFormalization}, we intend to describe a general polynomial proxy framework, calibrated at a given date $t$ ($t\geq1$ as in Vedani and Devineau \cCite{Vedani2013} for example, but also $t=0^{+}$ which is the case in the framework developed and implemented in Sections \ref{QuantitativeApproach} and \ref{LSMC_ContinuousCompliance}). We will then discriminate two distinct regressions ($LSMC$ and $CF$) that can be apprehended in a probabilistic or a statistic framework.

In a probabilistic framework, we note $NAV_{t}$ the Net Asset Value at date $t$, $NPV_{t}$ the Net present Value of margins variable seen at date $t$ and $\widehat{NAV}_{t}$ the estimated variable considered in the $CF$ regression. These variables are $\mathbf{P}\otimes \mathbf{Q}_{t}$ - measurable, denoting by $\mathbf{P} \otimes \mathbf{Q}_{t}$ the probability measure as introduced in Section 4 of Vedani and Devineau \cCite{Vedani2013}. The indexation by $\mathbf{P}\otimes\mathbf{Q}_t$ will be omitted for the sake of simplicity. Finally, we note $\mathcal{F}_{t}^{RW}$ the filtration that characterizes the real-world economic information contained between $0$ and $t$, $R_{u}$ the profit realized between dates $u-1$ and $u\geq1$, and $\delta_{u}$ the discount factor at date $u$.

We have,
\begin{center}
$NPV_{t}=\sum_{u=1}^{t+H}\frac{\delta_{u}}{\delta_{t}}R_{u}$
\end{center}

and denoting $NPV_{t}^{1},...,NPV_{t}^{P}$, $P$ \emph{i.i.d.} random variables conditionally to $\mathcal{F}_{t}^{RW}$, which follow the same probability distribution as $NPV_{t}$, conditionally to $\mathcal{F}_{t}^{RW}$,
\begin{center}
$\widehat{NAV}_{t}=\frac{1}{P}\sum_{p=1}^{P}NPV^{p}_{t}$,
\end{center}

for a chosen $P$ number of secondary scenarios. Moreover,
\begin{center}
$NAV_{t}=\mathbb{E}\left[\widehat{NAV}_{t}|\mathcal{F}_{t}^{MR}\right]=\mathbb{E}\left[NPV_{t}|\mathcal{F}_{t}^{MR}\right]$.
\end{center}

Denoting by $x_{t}$ the chosen regressors random vector (intercept included), ${}^{1}\!\beta$ (resp. ${}^{2}\!\beta$) the true value of the $CF$ (resp. $LSMC$) regression parameters, and $^{1}u_{t}$  (resp. $^{2}u_{t}$) the residual of the $CF$ (resp. $LSMC$) regression, both considered regressions can be written as follow. 
\begin{center}
$
\left\{
    \begin{array}{ll}
        (\text{regression}~1 - CF)~\widehat{NAV}_{t}=x_{t} {}^{1}\!\beta+^1 u_{t} & \mbox{under the assumption } \mathbb{E}\left[\widehat{NAV}_{t}|x_{t}\right]=x_{t} {}^{1}\!\beta \\
        (\text{regression}~2 - LSMC)~NPV_{t}=x_{t} {}^{2}\!\beta+^{2} u_{t} & \mbox{under the assumption } \mathbb{E}\left[NPV_{t}|x_{t}\right]=x_{t} {}^{2}\!\beta
		\end{array}
\right.
$
\end{center}

Note that this probabilist framework is the one chosen in Monfort \cCite{Monfort1988}.

Moreover, as seen in Vedani and Devineau \cCite{Vedani2013} and Kalberer \cCite{Kalberer2012}, we have ${}^{1}\!\beta={}^{2}\!\beta$ ($=\beta$, in the rest of this paper).

In a statistical framework one will first consider the samples used for the model calibration. As a consequence, let $^{1}N$ (resp. $^{2}N$) be the length of the calibration sample used in a $CF$ (resp. $LSMC$) regression, $\left({}^{1}\!x_{t}^{n}\right)_{n\in\llbracket1;^{1}N\rrbracket}$ (resp. $\left({}^{2}\!x_{t}^{n}\right)_{n\in\llbracket1;^{2}N\rrbracket}$) the $x_{t}$ outcomes, $\left(\widehat{NAV}_{t}^{n}\right)_{n\in\llbracket1;^{1}N\rrbracket}$ (resp. $\left(NPV_{t}^{n}\right)_{n\in\llbracket1;^{2}N\rrbracket}$) the associated $\widehat{NAV}_{t}$ (resp. $NPV_{t}$) outcomes\footnote{For a given value of $P$ for the simulation of the $\left(\widehat{NAV}_{t}^{n}\right)_{n\in\llbracket1;^{1}N\rrbracket}$ sample.} and $\left(^{1}u_{t}^{n}\right)_{n\in\llbracket1;^{1}N\rrbracket}$ (resp. $\left(^{2}u_{t}^{n}\right)_{n\in\llbracket1;^{2}N\rrbracket}$ the associated residuals. Note that in order to compare the relative efficiency of both approaches we will obviously have to consider an equal algorithmic complexity of the two approaches, which means $^{2}N=^{1}N \times P$.

In a statistical matrix framework we have,
\begin{center}
$
\left\{
    \begin{array}{ll}
        (\text{regression}~1 - CF)~ ^{1}Y_{t}={}^{1}\!X_{t} {}^{1}\!\beta+^1 U_{t} & \mbox{under the assumption } \mathbb{E}\left[^{1}Y_{t}|{}^{1}\!X_{t}\right]={}^{1}\!X_{t} {}^{1}\!\beta \\
        (\text{regression}~2 - LSMC)~ ^{2}Y_{t}={}^{2}\!X_{t} {}^{2}\!\beta+^{2} U_{t} & \mbox{under the assumption } \mathbb{E}\left[^{2}Y_{t}|{}^{2}\!X_{t}\right]={}^{2}\!X_{t} {}^{2}\!\beta
		\end{array}
\right.
$
\end{center}

Denoting,

\[^{1}Y_{t}=
\left(\begin{array}{c}
\widehat{NAV}^{1}_{t}\\
\vdots\\
\widehat{NAV}^{^{1}N}_{t}\\
\end{array}
\right)
\quad \text{and}
\quad
^{2}Y_{t}=
\left(\begin{array}{c}
NPV^{1}_{t}\\
\vdots\\
NPV^{^{2}N}_{t}\\
\end{array}
\right),\]

\[{}^{1}\!X_{t}=
\left(\begin{array}{c}
{}^{1}\!x^{1}_{t}\\
\vdots\\
{}^{1}\!x^{^{1}N}_{t}\\
\end{array}
\right)
\quad \text{and}
\quad
{}^{2}\!X_{t}=
\left(\begin{array}{c}
{}^{2}\!x^{1}_{t}\\
\vdots\\
{}^{2}\!x^{^{2}N}_{t}\\
\end{array}
\right),\]

\[^{1}U_{t}=
\left(\begin{array}{c}
^{1}u^{1}_{t}\\
\vdots\\
^{1}u^{^{1}N}_{t}\\
\end{array}
\right)
\quad \text{and}
\quad
^{2}U_{t}=
\left(\begin{array}{c}
^{2}u^{1}_{t}\\
\vdots\\
^{2}u^{^{2}N}_{t}\\
\end{array}
\right).\]

For example, this statistical framework is the one developed in Crépon and Jacquemet \cCite{Crepon2010}.
 
As the study goes forward and for the sake of simplicity, the time index will be omitted. 

\subsection{Comparison tools in a finite sample framework}
\label{FiniteSampleComparison}

In Subsection \ref{FiniteSampleComparison} we determine comparison elements to challenge the comparative efficiency of the $CF$ and $LSMC$ estimators, based on standard finite sample econometric results. We will see below that, in the general case, it is necessary to study first the properties of the residuals covariance matrices. Note that we will now consider the regressions in a statistical vision, more representative of the finite sample framework, and two assumptions will be made, which are verified in practice.
\begin{itemize}
	\item[] $\mathcal{H}_{0}:~\text{The}~\left(\widehat{NAV}^{i},{}^{1}\!x^{i}\right)~(\text{resp.}~(NPV^{i},{}^{2}\!x^{i}))~\text{outcomes are i.i.d.}$
	\item[] $\mathcal{H}_{1}:~\text{The matrix}~{}^{1}\!X'{}^{1}\!X~(\text{resp.}~{}^{2}\!X'{}^{2}\!X)~\text{is invertible.}$
\end{itemize}

Under these assumptions, the $OLS$ parameters estimators are respectively,
\begin{center}
${}^{1}\!\hat{\beta}=\left({}^{1}\!X'{}^{1}\!X\right)^{-1}\left({}^{1}\!X'^{1}Y\right)$ and ${}^{2}\!\hat{\beta}=\left({}^{2}\!X'{}^{2}\!X\right)^{-1}\left({}^{2}\!X'^{2}Y\right)$
\end{center}

These two estimators are consistent and unbiased.

In the following subsections we will introduce two comparison tools for the $LSMC$ and $CF$ methodologies in a finite sample framework: estimators of the parameters covariance matrices and asymptotic confidence intervals.

As far as the estimated covariance matrices are concerned, it is complicated to use them to compare models except when the eigenvalues of one matrix are all inferior to the eigenvalues of the other one. In this case the partial order on the hilbertian matrices tells us that the methodology leading to the first matrix is better (see Horn and Johnson \cCite{Horn2012} for more insight). However, this seldom happens in practice.  

As far as the asymptotic confidence intervals are concerned, we are able to compare the length of these intervals, obtained on the same set of primary scenarios, for the $\hat{\beta}$ estimated with the two different methodologies. If one methodology leads to smaller lengths than the other, it is the better one. This is the approach we will use in our empirical tests.

\subsubsection{Estimators covariance matrices under an homoskedasticity assumption}
\label{EstimatorsCovMatrices}

In Subsection \ref{EstimatorsCovMatrices} we add an homoskedasticity assumption for the residuals of both models,

\begin{center}
$\mathcal{H}_{2}:~\mathbb{V}\left[^{1}U|{}^{1}\!X\right]={}^{1}\!\sigma^{2}.I_{^{1}N}$ and $\mathbb{V}\left[^{2}U|{}^{2}\!X\right]={}^{2}\!\sigma^{2}.I_{^{2}N}$,
\end{center}

denoting by $I_{N}$ the identity matrix with rank $N$.

$\mathcal{H}_{2}$ can be operationally tested using an homoskedasticity test such as the Breusch and Pagan \cCite{Breusch1979}, the White \cCite{White1980} or the Goldfeld --- Quandt \cCite{Goldfeld1965} test. This assumption makes the Gauss --- Markov theorem applicable (see Plackett \cCite{Plackett1950}) and the $OLS$ estimators are the Best Linear Unbiased Estimators ($BLUE$). This means that considering the same calibration samples it is impossible to find less volatile estimators than the $OLS$ ones. Under this assumption it is also easy to assess the estimators' covariance matrices, conditionally to the explicative variables,
\begin{equation*}
\begin{split}
\mathbb{V}\left[{}^{1}\!\hat{\beta}|{}^{1}\!X\right] & = \left({}^{1}\!X'{}^{1}\!X\right)^{-1}\left({}^{1}\!X'\mathbb{V}\left[^{1}Y|{}^{1}\!X\right]{}^{1}\!X\right)\left({}^{1}\!X'{}^{1}\!X\right)^{-1}, \\
\mathbb{V}\left[{}^{1}\!\hat{\beta}|{}^{1}\!X\right] & = \left({}^{1}\!X'{}^{1}\!X\right)^{-1}\left({}^{1}\!X'\mathbb{V}\left[^{1}U|{}^{1}\!X\right]{}^{1}\!X\right)\left({}^{1}\!X'{}^{1}\!X\right)^{-1}, \\
\mathbb{V}\left[{}^{1}\!\hat{\beta}|{}^{1}\!X\right] & = {}^{1}\!\sigma^{2}\left({}^{1}\!X'{}^{1}\!X\right)^{-1}.
\end{split}
\end{equation*}

And, similarly, $\mathbb{V}\left[{}^{2}\!\hat{\beta}|{}^{2}\!X\right]={}^{2}\!\sigma^{2}\left({}^{2}\!X'{}^{2}\!X\right)^{-1}$.

Moreover, we can express consistent and unbiased estimators of ${}^{1}\!\sigma^{2}$ and ${}^{2}\!\sigma^{2}$. Let $K+1$ be the dimention of $x$, these estimators are respectively,
\begin{center}
$^{1}\hat{\sigma}^{2}=\frac{1}{^{1}N-K-1}\sum_{n=1}^{^{1}N}{}^{1}\!\hat{u}^{n^{2}}$ and $^{2}\hat{\sigma}^{2}=\frac{2}{^{2}N-K-1}\sum_{n=1}^{^{2}N}{}^{2}\!\hat{u}^{n^{2}}$,
\end{center}
with ${}^{1}\!\hat{u}^{n}=\widehat{NAV}^{n}-{}^{1}\!x^{n}{}^{1}\!\hat{\beta}$ and ${}^{2}\!\hat{u}^{n}=NPV^{n}-{}^{2}\!x^{n}{}^{2}\!\hat{\beta}$, the empirical residuals of regressions $1$ and $2$.

We therefore get two unbiased estimators of the previously given conditional covariance matrices,
\begin{center}
$\hat{\mathbb{V}}\left[{}^{1}\!\beta|{}^{1}\!X\right]=^{1}\hat{\sigma}^{2}\left({}^{1}\!X'{}^{1}\!X\right)^{-1}$ and $\hat{\mathbb{V}}\left[{}^{2}\!\beta|{}^{2}\!X\right]=^{2}\hat{\sigma}^{2}\left({}^{2}\!X'{}^{2}\!X\right)^{-1}$.
\end{center}

Eventually we have the two following convergences in distribution,
\begin{center}
$\mathbb{V}\left[{}^{1}\!\hat{\beta}|{}^{1}\!X\right]^{-\frac{1}{2}}\left({}^{1}\!\hat{\beta}-\beta\right)\stackrel{d}{\rightsquigarrow}\mathcal{N}\left(
\begin{pmatrix}
0\\
\vdots\\
0\\
\end{pmatrix},
I_{K+1}
\right)$\\
and $\mathbb{V}\left[{}^{2}\!\hat{\beta}|{}^{2}\!X\right]^{-\frac{1}{2}}\left({}^{2}\!\hat{\beta}-\beta\right)\stackrel{d}{\rightsquigarrow}\mathcal{N}\left(
\begin{pmatrix}
0\\
\vdots\\
0\\
\end{pmatrix},
I_{K+1}
\right)$.
\end{center}

Moreover, using the previously given estimators and adding simple assumptions on the first moments of the regressors (generally verified in practice), we have,
\begin{center}
$\hat{\mathbb{V}}\left[{}^{1}\!\hat{\beta}|{}^{1}\!X\right]^{-\frac{1}{2}}\left({}^{1}\!\hat{\beta}-\beta\right)\stackrel{d}{\rightsquigarrow}\mathcal{N}\left(
\begin{pmatrix}
0\\
\vdots\\
0\\
\end{pmatrix},
I_{K+1}
\right)$\\
and $\hat{\mathbb{V}}\left[{}^{2}\!\hat{\beta}|{}^{2}\!X\right]^{-\frac{1}{2}}\left({}^{2}\!\hat{\beta}-\beta\right)\stackrel{d}{\rightsquigarrow}\mathcal{N}\left(
\begin{pmatrix}
0\\
\vdots\\
0\\
\end{pmatrix},
I_{K+1}
\right)$.\
\end{center}

\subsubsection{Comparison between the estimators covariance matrices without the homoskedasticity assumption}

In practice it is unusual to observe homoskedastic residuals. We now suppress $\mathcal{H}_{2}$ in order to consider a more robust framework. Note first that, in the heteroskedastic case, the $OLS$ estimators are no longer the $BLUE$.

Moreover, in this new framework we do not have a simple form for the estimators' covariance matrices any more,
\begin{equation*}
\begin{split}
	\mathbb{V}\left[{}^{1}\!\hat{\beta}|{}^{1}\!X\right] & = \left({}^{1}\!X'{}^{1}\!X\right)^{-1}\left({}^{1}\!X'\mathbb{V}\left[{}^{1}\!U|{}^{1}\!X\right]{}^{1}\!X\right)\left({}^{1}\!X'{}^{1}\!X\right)^{-1}\\
\text{and} \mathbb{V}\left[{}^{2}\!\hat{\beta}|{}^{2}\!X\right] & = \left({}^{2}\!X'{}^{2}\!X\right)^{-1}\left({}^{2}\!X'\mathbb{V}\left[{}^{2}\!U|{}^{2}\!X\right]{}^{2}\!X\right)\left({}^{2}\!X'{}^{1}\!X\right)^{-1}.
\end{split}
\end{equation*}

However, we still have the two following convergences in distribution,
\begin{center}
$\mathbb{V}\left[{}^{1}\!\hat{\beta}|{}^{1}\!X\right]^{\frac{1}{2}}\left({}^{1}\!\hat{\beta}-\beta\right)\stackrel{d}{\rightsquigarrow}\mathcal{N}\left(
\begin{pmatrix}
0\\
\vdots\\
0\\
\end{pmatrix},
I_{K+1}
\right)$\\
and $\mathbb{V}\left[{}^{2}\!\hat{\beta}|{}^{2}\!X\right]^{-\frac{1}{2}}\left({}^{2}\!\hat{\beta}-\beta\right)\stackrel{d}{\rightsquigarrow}\mathcal{N}\left(
\begin{pmatrix}
0\\
\vdots\\
0\\
\end{pmatrix},
I_{K+1}
\right)$
\end{center}

White \cCite{White1980} proposes the use of a biased estimator of the residuals variance, in the case of independent calibration samples. In our case, it aims at resorting to the following estimators,\
\begin{center}
$\hat{\mathbb{V}}\left[{}^{1}\!U|{}^{1}X\right]=
\begin{pmatrix}
{}^{1}\!\hat{u}^{1^{2}} & \cdots & 0\\
\vdots & \ddots & \vdots\\
0 & \cdots & {}^{1}\!\hat{u}^{{}^{1}\!N^{{}^{2}}}\\
\end{pmatrix}$ and $
\hat{\mathbb{V}}\left[{}^{1}\!U|{}^{1}X\right]=
\begin{pmatrix}
{}^{2}\!\hat{u}^{1^{2}} & \cdots & 0\\
\vdots & \ddots & \vdots\\
0 & \cdots & {}^{2}\!\hat{u}^{{}^{2}\!N^{{}^{2}}}\\
\end{pmatrix}$.
\end{center}

Note that other \emph{less biased} estimators are proposed in MacKinnon and White \cCite{Mackinnon1985}. These adapted estimators are less used in practice and will not be considered in this paper. This new data enables one to assess two biased but consistent estimators of the covariance matrices of the $OLS$ estimators,
\begin{equation*}
\begin{split}
	\hat{\mathbb{V}}^{White}\left[{}^{1}\!\hat{\beta}|{}^{1}\!X\right] & =\left({}^{1}\!X'{}^{1}\!X\right)^{-1}\left(\sum_{n=1}^{{}^{1}\!N}{}^{1}\!\hat{u}^{n^{2}}{}^{1}\!x^{n'}{}^{1}\!x^{n}\right)\left({}^{1}\!X'{}^{1}\!X\right)^{-1}\\
\text{and} \hat{\mathbb{V}}^{White}\left[{}^{2}\!\hat{\beta}|{}^{2}\!X\right]& =\left({}^{2}\!X'{}^{2}\!X\right)^{-1}\left(\sum_{n=1}^{{}^{2}\!N}{}^{2}\!\hat{u}^{n^{2}}{}^{2}\!x^{n'}{}^{2}\!x^{n}\right)\left({}^{2}\!X'{}^{2}\!X\right)^{-1}
\end{split}
\end{equation*}

Moreover, under simple assumptions concerning the first moments of the regressors (generally verified in practice), these estimators enables one to obtain the following convergences in distribution,\
\begin{center}
$\hat{\mathbb{V}}^{White}\left[{}^{1}\!\hat{\beta}|{}^{1}\!X\right]^{-\frac{1}{2}}\left({}^{1}\!\hat{\beta}-\beta\right)\stackrel{d}{\rightsquigarrow}\mathcal{N}\left(
\begin{pmatrix}
0\\
\vdots\\
0\\
\end{pmatrix},
I_{K+1}
\right)$\\
and $\hat{\mathbb{V}}^{White}\left[{}^{2}\!\hat{\beta}|{}^{2}\!X\right]^{-\frac{1}{2}}\left({}^{2}\!\hat{\beta}-\beta\right)\stackrel{d}{\rightsquigarrow}\mathcal{N}\left(
\begin{pmatrix}
0\\
\vdots\\
0\\
\end{pmatrix},
I_{K+1}
\right)$
\end{center}

To conclude on the heteroskedastic framework, it is important to note that these variance-covariance matrices estimators are biased and sometimes very volatile. The heteroskedastic framework is more general and robust than the homoskedastic framework. It is generally more adapted to our proxy methodologies.

\subsection{Asymptotic confidence intervals}

In practice, the length of the asymptotic confidence intervals given the estimators of the covariance matrices are good comparison tools provided by both homoskedastic and heteroskedastic frameworks. This subsection describes the construction steps of these intervals.

\subsubsection{Asymptotic confidence intervals under the homoskedasticity assumption}

If $\mathcal{H}_{2}$ is assumed, an asymptotic confidence interval for the approximate $NAV$ can be obtained, using the following convergences in law,\
\begin{center}
$\hat{\mathbb{V}}\left[{}^{1}\!\hat{\beta}|{}^{1}\!X\right]^{-\frac{1}{2}}\left({}^{1}\!\hat{\beta}-\beta\right)\stackrel{d}{\rightsquigarrow}\mathcal{N}\left(
\begin{pmatrix}
0\\
\vdots\\
0\\
\end{pmatrix},
I_{K+1}
\right)$\\
and $\hat{\mathbb{V}}\left[{}^{2}\!\hat{\beta}|{}^{2}\!X\right]^{-\frac{1}{2}}\left({}^{2}\!\hat{\beta}-\beta\right)\stackrel{d}{\rightsquigarrow}\mathcal{N}\left(
\begin{pmatrix}
0\\
\vdots\\
0\\
\end{pmatrix},
I_{K+1}
\right)$.\
\end{center}

For the $CF$ regression, the $\alpha\%$ (with $\alpha\%$ close to $1$) asymptotic confidence interval obtained from this formula, for $\bar{x}$, a given regressors' outcome, is,
\begin{center}
${}^{1}\!IC^{{}^{1}\!N}_{\alpha\%}\left(\bar{x}\beta\right)=\Biggl[\bar{x}{}^{1}\!\hat{\beta}\pm\!q_{\frac{1+\alpha\%}{2}}\sqrt{{}^{1}\!\hat{\sigma}^{2}\left(\bar{x}\left({}^{1}\!X'{}^{1}\!X\right)^{-1}\bar{x}'\right)}\Biggr]$,
\end{center}
and for the $LSMC$,
\begin{center}
${}^{2}\!IC^{{}^{2}\!N}_{\alpha\%}\left(\bar{x}\beta\right)=\Biggl[\bar{x}{}^{2}\!\hat{\beta}\pm\!q_{\frac{1+\alpha\%}{2}}\sqrt{{}^{2}\!\hat{\sigma}^{2}\left(\bar{x}\left({}^{2}\!X'{}^{2}\!X\right)^{-1}\bar{x}'\right)}\Biggr]$,\\
\end{center}
denoting by $q_{\frac{1+\alpha\%}{2}}$ the $\frac{1+\alpha\%}{2}$ quantile of a standard Gaussian distribution.

\subsubsection{Asymptotic confidence intervals without the homoskedasticity assumption}

Without the homoskedasticity assumption it is also possible to build asymptotic confidence intervals, based on White’s estimator \cCite{White1980} properties. Indeed, this estimator enables one to assess the following convergence in law,\
\begin{center}
$\hat{\mathbb{V}}^{White}\left[{}^{1}\!\hat{\beta}|{}^{1}\!X\right]^{-\frac{1}{2}}\left({}^{1}\!\hat{\beta}-\beta\right)\stackrel{d}{\rightsquigarrow}\mathcal{N}\left(
\begin{pmatrix}
0\\
\vdots\\
0\\
\end{pmatrix},
I_{K+1}
\right)$\\
and $\hat{\mathbb{V}}^{White}\left[{}^{2}\!\hat{\beta}|{}^{2}\!X\right]^{-\frac{1}{2}}\left({}^{2}\!\hat{\beta}-\beta\right)\stackrel{d}{\rightsquigarrow}\mathcal{N}\left(
\begin{pmatrix}
0\\
\vdots\\
0\\
\end{pmatrix},
I_{K+1}
\right)$
\end{center}

For the $CF$ regression, the $\alpha\%$ (with $\alpha\%$ close to $1$) asymptotic confidence interval obtained from this formula, for $\bar{x}$, a given regressors' outcome, is,
\begin{center}
${}^{1}\!IC^{{}^{1}\!N}_{\alpha\%}\left(\bar{x}\beta\right)=\Biggl[\bar{x}{}^{1}\!\hat{\beta}\pm\!q_{\frac{1+\alpha\%}{2}}\sqrt{\bar{x}\left({}^{1}\!X'{}^{1}\!X\right)^{-1}\left(\sum_{n=1}^{{}^{1}\!N}{}^{1}\!\hat{u}^{n^{2}}{}^{1}\!x^{n'}{}^{1}\!x^{n}\right)\left({}^{1}\!X'{}^{1}\!X\right)^{-1}\bar{x}'}\Biggr]$,
\end{center}
and for the $LSMC$,\ 
\begin{center}
${}^{2}\!IC^{{}^{2}\!N}_{\alpha\%}\left(\bar{x}\beta\right)=\Biggl[\bar{x}{}^{2}\!\hat{\beta}\pm\!q_{\frac{1+\alpha\%}{2}}\sqrt{\bar{x}\left({}^{2}\!X'{}^{2}\!X\right)^{-1}\left(\sum_{n=1}^{{}^{2}\!N}{}^{2}\!\hat{u}^{n^{2}}{}^{2}\!x^{n'}{}^{2}\!x^{n}\right)\left({}^{2}\!X'{}^{2}\!X\right)^{-1}\bar{x}'}\Biggr]$.\\
\end{center}

Consider now a set of $N$ independent outcomes following the same distribution as $x$, $\left(\bar{x}^{i}\right)_{i\in\llbracket1;N\rrbracket}$. It is possible to calculate the lengths of the asymptotic confidence intervals built for both $CF$ and $LSMC$, and to compare these values to assess which estimator is more efficient in practice (this will be used to see what happens when the number of risk factors increases).

In the following subsection we will test empirically the results previously presented when $t=0^{+}$ (continuous compliance framework).

\subsection{Empirical tests}
\label{Implementation}

\subsubsection{Implementation framework}

The implementation framework used in this section is the same as the one presented in Section \ref{LSMC_ContinuousCompliance}.

The $LSMC$ approach has been run on a sample of 50~000 independent $NPV_{0^{+}}$ outcomes. To equalize the algorithmic complexity between both procedures, the $CF$ approach has been launched on a sample of $100$ independent $\widehat{NAV}_{0^{+}}$ outcomes, calculated as means of $500~NPV_{0^{+}}$ ($100$ primary scenarios $\times$ 500 secondary scenarios).

To consider a more statistically efficient (due to the larger number of outcomes) implementation framework we have chosen the $LSMC$ methodology as a base to assess the optimal set of regressors. 

For each given number of risk factors $J (J=1,...,4)$, these have been designated using a stepwise backward approach based on the $AIC$ stopping criteria and on an initialization set of potential regressors given by,
\begin{center}
$\left\{{}^{i}\!\varepsilon^{k}.{}^{j}\!\varepsilon^{l}, \forall i,j\in\llbracket1;J\rrbracket,\forall\!k,l\in \mathbb{N} | k+l\leq3\right\}$, 
\end{center}
denoting by ${}^{i}\!\varepsilon^{k}$ the $i$-th risk factor power $k$. 

The implementation steps are the same for each value of $J$,
\begin{itemize}
	\item	assessment of the $LSMC$ optimal set of regressors ${}^{J}\!x$ and $OLS$ estimator ${}^{J}\!\hat{\beta}^{LSMC}$,
	\item use of the ${}^{J}\!x$ set of regressors to obtain the associated $CF$ $OLS$ estimator ${}^{J}\hat{\beta}^{CF}$,
	\item implementation of a Breusch-Pagan homoskedasticity test on the $LSMC$ methodology (there are too few outcomes to use a Breusch-Pagan test on the $CF$ approach),
	\item comparison of the confidence interval lengths obtained on the 50~000 primary scenarios sample used to implement the $LSMC$ approach.
\end{itemize}

\subsubsection{Heteroskedasticity test}

In this study the heteroskedasticity has been tested using a Breusch-Pagan test. The following results have been obtained on the various $LSMC$ models on one, two, three and four risk factors data.
\begin{table}[h!]
\centering
\caption{Breusch-Pagan tests --- $LSMC$ data.}
\begin{tabular}{|l|c|c|c|c|}
	\hline
	 \multicolumn{1}{|c|}{\textbf{$LSMC$ methodology}} & \multicolumn{1}{c|}{\textbf{1 risk factor}} & \multicolumn{1}{c|}{\textbf{2 risk factor}} & \multicolumn{1}{c|}{\textbf{3 risk factor}} & \multicolumn{1}{c|}{\textbf{4 risk factor}}\\
 \hline
	 Breusch-Pagan statistic & 25.0 & 41.6 & 50.7 & 76.0\\
	 Breusch-Pagan p-value & 5.8e-07 & 7.3e-08 & 2.2e-06 & 4.5e-06\\
	\hline
\end{tabular}
\end{table}

The tests, and the homoskedastic assumption, are rejected even for a significance level of $1\%$. Note that there are too few $CF$ implementation data ($100$ outcomes per number of risk factor) to assess robust homoskedasticity tests.

In the following subsections we will study the results obtained with both the $LSMC$ and $CF$ methodologies (1/2/3/4 risk factors), for both the heteroskedastic and homoskedastic formulas.

\subsubsection{Results in the homoskedastic framework}

Turning from an homoskedastic to an heteroskedastic framework enables to obtain more robust results. Moreover, the heteroskedastic scheme seems more adapted for our study. However, the homoskedastic formulas provide interesting results that can be compared to those obtained using the heteroskedastic formulas in order to conclude on this empirical subsection. The comparison of the homoskedastic parameters covariance matrix estimators provides the following results.

\paragraph{One risk factor framework.}

Only two significant regressors have been selected after implementing a backward stepwise methodology with an $AIC$ stopping criteria.
\begin{table}[h!]
\centering
\caption{$LSMC$ covariance matrix eigenvalues --- 1 risk factor (stock).}
\begin{tabular}{|l|r|r|}
	\hline
	\multicolumn{1}{|c|}{\textbf{1 risk factor}} & \multicolumn{1}{c|}{\textbf{$\lambda_{1}$}} & \multicolumn{1}{c|}{\textbf{$\lambda_{2}$}} \\
 \hline
	 $LSMC$ & 5.21e+15 & 2.69e+14\\
	 $CF$ & 5.80e+15 & 2.66e+14\\
	\hline
\end{tabular}
\end{table}

Here below we display the tables comparing the asymptotic confidence intervals' lengths on the $50~000$ primary scenarios used in the $LSMC$ implementation.

\begin{table}[h!]
\centering
\caption{Asymptotic confidence intervals lengths --- 1 risk factor (stock).}
\begin{tabular}{| l l | C{\tabularSmallColWidth} | C{\tabularSmallColWidth} |}
	\hline
	\multicolumn{2}{|c|}{\textbf{1 risk factor}} & \multicolumn{1}{c|}{\textbf{$LSMC$}} & \multicolumn{1}{c|}{\textbf{$CF$}} \tabularnewline
 \hline
	\multicolumn{2}{|l|}{Number of smaller asymptotic confidence} & 38~956 & 11~044 \tabularnewline
	\multicolumn{2}{|l|}{intervals (max = 50~000 scenarios)} & ($77.9\%$) & ($22.1\%$) \tabularnewline
	\hline
\end{tabular}
\end{table}

On average (on the 50~000 scenarios), the $LSMC$ methodology leads to a slightly smaller asymptotic confidence interval than the $CF$. Moreover, the $LSMC$ approach leads to a lesser asymptotic confidence interval for $77.9\%$ of the 50~000 independent scenarios considered here.

\paragraph{Two risk factor framework.}

Six significant regressors are selected after implementing a backward stepwise methodology with an $AIC$ stopping criteria.
\begin{table}[h!]
\centering
\caption{LSMC covariance matrix eigenvalues --- 2 risk factors (stock, interest rates).}
\begin{tabular}{|l|c|c|c|c|c|c|}
	\hline
	\multicolumn{1}{|c|}{\textbf{2 risk factors}} & \multicolumn{1}{c|}{\textbf{$\lambda_{1}$}} & \multicolumn{1}{c|}{\textbf{$\lambda_{2}$}} & \multicolumn{1}{c|}{\textbf{$\lambda_{3}$}} & \multicolumn{1}{c|}{\textbf{$\lambda_{4}$}} & \multicolumn{1}{c|}{\textbf{$\lambda_{5}$}} & \multicolumn{1}{c|}{\textbf{$\lambda_{6}$}} \\
 \hline
	\hline
	 $LSMC$ & 3.64e+18 & 6.02e+16 & 2.27e+16 & 4.77e+15 & 1.88e+15 & 2.08e+14\\
	 $CF$ & 3.64e+15 & 6.87e+14 & 2.18e+16 & 5.00e+15 & 1.87e+15 & 1.98e+14\\
	\hline
\end{tabular}
\end{table}

Here below we display the tables comparing the asymptotic confidence intervals' lengths on the $50~000$ primary scenarios used in the $LSMC$ implementation.

\begin{table}[h!]
\centering
\caption{Asymptotic confidence intervals lengths --- 2 risk factors (stock, interest rates).}
\begin{tabular}{| l l | C{\tabularSmallColWidth} | C{\tabularSmallColWidth} |}
	\hline
	 \multicolumn{2}{|c|}{\textbf{2 risk factors}} & \multicolumn{1}{c|}{\textbf{$LSMC$}} & \multicolumn{1}{c|}{\textbf{$CF$}} \tabularnewline
 \hline
	\multicolumn{2}{|l|}{Number of smaller asymptotic confidence} & 22~228 & 27~772 \tabularnewline
	\multicolumn{2}{|l|}{intervals (max = 50~000 scenarios)} & ($44.5\%$) & ($55.5\%$) \tabularnewline
	\hline
\end{tabular}
\end{table}

\paragraph{Three risk factor framework.}

Fourteen significant regressors are selected after implementing a backward stepwise methodology with an $AIC$ stopping criteria. 

Here below we display the tables comparing the asymptotic confidence intervals' lengths on the $50~000$ primary scenarios used in the $LSMC$ implementation.\footnote{The eigenvalues of the covariance matrix estimator, only presented as illustrations for the 1 and 2 risk factors frameworks, are omitted, for the sake of simplicity, in the following studied cases.}.

\begin{table}[h!]
\centering
\caption{Asymptotic confidence intervals lengths --- 3 risk factors (stock, IR, corporate spread).}
\begin{tabular}{| l l | C{\tabularSmallColWidth} | C{\tabularSmallColWidth} |}
	\hline
	 \multicolumn{2}{|c|}{\textbf{3 risk factors}} & \multicolumn{1}{c|}{\textbf{$LSMC$}} & \multicolumn{1}{c|}{\textbf{$CF$}} \tabularnewline
 \hline
	\multicolumn{2}{|l|}{Number of smaller asymptotic confidence} & 45~398 & 4~602 \tabularnewline
	\multicolumn{2}{|l|}{intervals (max = 50~000 scenarios)} & ($90.8\%$) & ($9.2\%$) \tabularnewline
	\hline
\end{tabular}
\end{table}

\paragraph{Four risk factor framework.}

Thirty significant regressors are selected after implementing a backward stepwise methodology with an $AIC$ stopping criteria. 

Here below we display the tables comparing the asymptotic confidence intervals' lengths on the $50~000$ primary scenarios used in the $LSMC$ implementation.

\begin{table}[h!]
\centering
\caption{Asymptotic confidence intervals lengths --- 4 risk factors (stock, IR, corporate spread, sovereign spread).}
\begin{tabular}{| l l | C{\tabularSmallColWidth} | C{\tabularSmallColWidth} |}
	\hline
	 \multicolumn{2}{|c|}{\textbf{4 risk factors}} & \multicolumn{1}{c|}{\textbf{$LSMC$}} & \multicolumn{1}{c|}{\textbf{$CF$}} \tabularnewline
 \hline
	\multicolumn{2}{|l|}{Number of smaller asymptotic confidence} & 42~044 & 7~956 \tabularnewline
	\multicolumn{2}{|l|}{intervals (max = 50~000 scenarios)} & ($84.1\%$) & ($15.9\%$) \tabularnewline
	\hline
\end{tabular}
\end{table}

We now present the same results obtained without the homoskedasticity assumptions.

\subsubsection{Results in the heteroskedastic framework}

Comparison of the homoskedastic parameters covariance matrix estimators.

\begin{table}[h!]
\centering
\caption{Asymptotic confidence intervals lengths --- 1 risk factor (stock).}
\begin{tabular}{| l l | C{\tabularSmallColWidth} | C{\tabularSmallColWidth} |}
	\hline
	 \multicolumn{2}{|c|}{\textbf{1 risk factor}} & \multicolumn{1}{c|}{\textbf{$LSMC$}} & \multicolumn{1}{c|}{\textbf{$CF$}} \tabularnewline
 \hline
	\multicolumn{2}{|l|}{Number of smaller asymptotic confidence} & 22~664 & 27~336 \tabularnewline
	\multicolumn{2}{|l|}{intervals (max = 50~000 scenarios)} & ($45.3\%$) & ($54.7\%$) \tabularnewline
	\hline
\end{tabular}
\end{table}

\begin{table}[h!]
\centering
\caption{Asymptotic confidence intervals lengths --- 2 risk factors (stock, interest rates).}
\begin{tabular}{| l l | C{\tabularSmallColWidth} | C{\tabularSmallColWidth} |}
	\hline
	 \multicolumn{2}{|c|}{\textbf{2 risk factors}} & \multicolumn{1}{c|}{\textbf{$LSMC$}} & \multicolumn{1}{c|}{\textbf{$CF$}} \tabularnewline
 \hline
	\multicolumn{2}{|l|}{Number of smaller asymptotic confidence} & 10~656 & 39~344 \tabularnewline
	\multicolumn{2}{|l|}{intervals (max = 50~000 scenarios)} & ($21.3\%$) & ($78.7\%$) \tabularnewline
	\hline
\end{tabular}
\end{table}

\begin{table}[h!]
\centering
\caption{Asymptotic confidence intervals lengths --- 3 risk factors (stock, IR, corporate spread).}
\begin{tabular}{| l l | C{\tabularSmallColWidth} | C{\tabularSmallColWidth} |}
	\hline
	 \multicolumn{2}{|c|}{\textbf{3 risk factors}} & \multicolumn{1}{c|}{\textbf{$LSMC$}} & \multicolumn{1}{c|}{\textbf{$CF$}} \tabularnewline
 \hline
	\multicolumn{2}{|l|}{Number of smaller asymptotic confidence} & 18~803 & 31~197 \tabularnewline
	\multicolumn{2}{|l|}{intervals (max = 50~000 scenarios)} & ($37.6\%$) & ($62.4\%$) \tabularnewline
	\hline
\end{tabular}
\end{table}

\begin{table}[h!]
\centering
\caption{Asymptotic confidence intervals lengths --- 4 risk factors (stock, IR, corporate spread, sovereign spread).}
\begin{tabular}{| l l | C{\tabularSmallColWidth} | C{\tabularSmallColWidth} |}
	\hline
	 \multicolumn{2}{|c|}{\textbf{4 risk factors}} & \multicolumn{1}{c|}{\textbf{$LSMC$}} & \multicolumn{1}{c|}{\textbf{$CF$}} \\
 \hline
	\multicolumn{2}{|l|}{Number of smaller asymptotic confidence} & 17~513 & 32~487 \tabularnewline
	\multicolumn{2}{|l|}{intervals (max = 50~000 scenarios)} & ($35.0\%$) & ($65.0\%$) \tabularnewline
	\hline
\end{tabular}
\end{table}\

\subsubsection{Conclusion on the empirical tests}

Two major comments can be made after having studied these results.

First, it is important to go further than just considering the results obtained under the homoskedasticity assumption. If these results alone are observed the $LSMC$ approach seems to be the best methodology in most cases. This is not the case when one studies the results provided without the homoskedasticity assumption. Note that the heteroskedastic framework is more robust in general and is much more realistic here, considering the Breusch-Pagan tests.

Second, we can assume that the heteroskedasticity shape has a great impact on the efficiency comparison between both $CF$ and $LSMC$ methodology in a finite sample framework. In particular, this directly modifies the $\hat{\mathbb{V}}^{White}$ estimator. One should note that there are several econometric methods to reduce the heteroskedasticity of our model that have not been tested here. For more insight on these approaches see Greene \cCite{Greene2003}.

In any case, our study does not evidence any superiority of the $LSMC$ over the $CF$ methodology. However, we have clearly seen, throughout the implementation, that the small number of outcomes considered in the $CF$ approach leads to statistical issues while assessing homoskedasticity tests and confidence intervals. The problem here seems to comes from the fact that there are too few outcomes to get through the sample bias embedded within the secondary scenarios tables used to calculate the $\widehat{NAV}_{0^{+}}$ outcomes. In opposition, the sample bias that comes from the $LSMC$ scenarios is mitigated between the primary simulations. Eventually, the squared errors of the $CF$ implementation are lower than they would be if calculated on more outcomes, which leads to artificially small confidence intervals. It is clear that this phenomenon takes more and more importance as the number of risk factors / regressors rises.

To conclude we can only advise practitioners to prefer an $LSMC$ methodology to assess approximate $NAV$ outcomes. The heteroskedasticity tests may always lead to a rejection of the homoskedasticity assumption but the confidence intervals obtained will always be more robust than those of a $CF$ approach. 

Note that this implementation and its conclusions correspond to a specific (but realistic) empirical framework. The authors did not aim at drawing general conclusions on the use of parametric proxies for life insurance $NAV$ projections. This section, initially only aiming at challenging the generally agreed idea that the $LSMC$ methodology is more robust than $CF$ in a large dimension context, is eventually a good opportunity to raise proxies implementation issues such as the heteroskedasticity management and the asymptotic confidence intervals assessment.

The authors notice that Subsection \ref{Implementation} could have been completed with the comparison the $CF$ and $LEMC$ results to real values of $NAV$. However, the real $NAV$ outcomes are unobservable in practice and good estimators imply a great number of secondary scenarios. We wanted here to stay in a practical scheme, with great algorithmic constraints.

\section{Conclusion}

The continuous compliance requirement is a key strategic issue for European insurers. In this article, we have presented the various characteristics of this problem and provided a monitoring tool to answer it. Our monitoring scheme is based on the implementation of parametric proxies, already used among the insurance players to project the Net Asset Value over time, adapted to fit the ORSA continuous compliance requirements. The tool has been implemented on a realistic life insurance portfolio to present the main features of both the development and the practical use of the monitoring scheme. In particular several other relevant possible uses for the tool are presented. In the last Section we have seen that the comparison of the Curve Fitting and the Least Squares Monte Carlo methodologies in a finite sample framework and considering an increasing dimensioning scheme, did not lead to firm conclusions but that the Least Squares Monte Carlo led to fewer statistical issues especially to assess robust asymptotic confidence intervals. In addition, this section has been an opportunity to raise several practical issues, such as the heteroskedasticity management and asymptotic confidence intervals calculation, concerning the use of polynomial proxies (both Least Squares Monte Carlo and Curve Fitting), in our framework (a life insurance savings product).

Note that the monitoring tool only provides approximate values and is based on assumptions that can be discussed. The authors notice that the modeling choices can lead to errors. In particular we can only advice the future users of our tool to update the proxies frequently in order to make sure that the underlying stability assumptions are reasonable. One of the future axes to investigate is clearly to aim at a better control of the error and to address in depth the issue of the proxies recalibration frequency. 

In addition the possibility to add asset-mix weights in the monitored risk factors set should be tested. This would greatly help asset managers to select optimal asset-mixes, consistently with the risk strategy of the undertaking.

Eventually we intend to investigate the issue of the calculation error due to the use of proxies in Life insurance projections and the various possibilities provided by the econometric theory to optimize the proxies calibration process, in order to decrease the heteroskedasticity of our models and the volatility of the obtained estimators. 

\section*{Acknowledgement}
\addcontentsline{toc}{section}{Acknowledgement}

The authors would like to address very special thanks to Fabien Conneau, Laurent Devineau and Christophe Vallet for their help all along the redaction of this paper. We would also like to thank St\'ephane Loisel for his relevant comments throughout the final review of the article. 

Moreover, we would like to extend our thanks to all the employees of Milliman Paris, and in particular the members of the R\&D team.

\def\refname{References}
\bibliographystyle{plainnat}
\bibliography{biblio}
\addcontentsline{toc}{section}{References}

\end{document}